\documentclass[twocolumn,nofootinbib,superscriptaddress]{revtex4-2}
\usepackage{hyperref}
\usepackage{bm}
\usepackage{graphicx}
\usepackage{braket}
\usepackage{aas_macros}
\usepackage{amsmath}
\usepackage{amssymb}
\usepackage{color}
\newcommand{\pccm}{{\rm pc}\,{\rm cm}^{-3}}

\begin{document}
%%%%%%%%%%%%%%%%%%%%%%%%%%%%%%%%%%%%%%%%%%
\title{Relativistic imprints on dispersion measure space distortions}
%%%%%%%%%%%%%%%%%%%%%%%%%%%%%%%%%%%%%%%%%%
\author{Shohei Saga}
\email{shohei.saga@yukawa.kyoto-u.ac.jp}
\affiliation{Institute for Advanced Research, Nagoya University, Furo-cho Chikusa-ku, Nagoya 464-8601, Japan}
\affiliation{Kobayashi-Maskawa Institute for the Origin of Particles and the
Universe, Nagoya University, Chikusa-ku, Nagoya, 464-8602, Japan}
\affiliation{Sorbonne Universit\'e, CNRS, UMR7095, Institut d'Astrophysique de Paris, 98bis boulevard Arago, F-75014 Paris, France}
\author{David Alonso}
\affiliation{Department of Physics, University of Oxford, Denys Wilkinson Building, Keble Road, Oxford OX1 3RH, United Kingdom}
%%%%%%%%%%%%%%%%%%%%%%%%%%%%%%%%%%%%%%%%%%
\date{\today}
%%%%%%%%%%%%%%%%%%%%%%%%%%%%%%%%%%%%%%%%%%
\begin{abstract}
  We investigate the three-dimensional clustering of sources emitting electromagnetic pulses traveling through cold electron plasma, whose radial distance is inferred from their dispersion measure. As a distance indicator, dispersion measure is systematically affected by inhomogeneities in the electron density along the line of sight and special and general relativistic effects, similar to the case of redshift surveys. We present analytic expressions for the correlation function of fast radio bursts (FRBs), and for the galaxy-FRB cross-correlation function, in the presence of these {\it dispersion measure-space distortions}. We find that the even multipoles of these correlations are primarily dominated by non-local contributions (e.g. the electron density fluctuations integrated along the line of sight), while the dipole also receives a significant contribution from the Doppler effect, one of the major relativistic effects. A large number of FRBs, $\mathcal{O}(10^{5}\sim10^{6})$, expected to be observed in the Square Kilometre Array, would be enough to measure the even multipoles at very high significance, ${\rm S}/{\rm N} \approx 100$, and perhaps to make a first detection of the dipole (${\rm S}/{\rm N} \approx 10$) in the FRB correlation function and FRB-galaxy cross correlation function. This measurement could open a new window to study and test cosmological models.
\end{abstract}
%%%%%%%%%%%%%%%%%%%%%%%%%%%%%%%%%%%%%%%%%%
\maketitle
%\tableofcontents
%%%%%%%%%%%%%%%%%%%%%%%%%%%%%%%%%%%%%%%%%%

%%%%%%%%%%%%%%%%%%%%%%%%%%%%%%%%%%%%%%%%%%
%%%%%%%%%%%%%%%%%%%%%%%%%%%%%%%%%%%%%%%%%%
%%%%%%%%%%%%%%%%%%%%%%%%%%%%%%%%%%%%%%%%%%
\section{Introduction}
%%%%%%%%%%%%%%%%%%%%%%%%%%%%%%%%%%%%%%%%%%
%%%%%%%%%%%%%%%%%%%%%%%%%%%%%%%%%%%%%%%%%%
%%%%%%%%%%%%%%%%%%%%%%%%%%%%%%%%%%%%%%%%%%

  Fast radio bursts (FRBs) are radio transients with millisecond duration first detected by Ref.~\cite{2007Sci...318..777L}. The radio pulses emitted from FRBs travel through the cold ionized intergalactic medium, and their arrival time receives a frequency dependence that depends on the column density of free electrons through the so-called dispersion measure (DM, see \cite{2020arXiv200702886K} for details). Recent observations of FRBs have found DMs far larger ($\gtrsim 500\,\pccm$) than would be expected for any single source ($\sim 100\,\pccm$), suggesting that they reside in extragalactic sources. This is supported by the redshifts measured for small samples of localised FRBs, although their specific origin is still under debate (see Refs.~\cite{2016PASA...33...45P,2017ApJ...834L...7T,2017Natur.541...58C,2020Natur.581..391M,2021ApJS..257...59C} and also Refs.~\cite{2019PhR...821....1P,2019ARA&A..57..417C,2019A&ARv..27....4P} for recent reviews). Future surveys such as UTMOST\footnote{\url{https://astronomy.swin.edu.au/research/utmost/}}, HIRAX\footnote{\url{https://hirax.ukzn.ac.za/}}, and Square Kilometre Array (SKA)\footnote{\url{https://www.skao.int/}} will detect more than thousands of FRBs per year~\cite{2016MNRAS.458..718C,2016SPIE.9906E..5XN,2016MNRAS.455.2207R,2016MNRAS.460.1054C,2020MNRAS.497.4107H}, and provide large FRB catalogs across cosmological scales. It is therefore interesting to study the possibility of using FRBs as a new cosmological probe, complementary to the cosmic microwave background and galaxy surveys.

  Since the DM is given by the column density of free electrons along the pulse's path, DM measurements may contain important information about the baryonic content of the intergalactic medium, such as, the reionization history~\cite{2014ApJ...783L..35D,2016JCAP...05..004F,2019arXiv190102418M,2019MNRAS.485.2281C,2021MNRAS.502.5134B,2021PhRvD.103j3526B,2021MNRAS.505.2195P,2022ApJ...933...57H}, the gas mass fraction in the cosmic web \cite{2019ApJ...876..146L,2019JCAP...09..039W,2020MNRAS.496L..28L,2023EPJC...83..138L}, baryonic effects \cite{2022JCAP...04..046N,2024arXiv240308611T}, and even important clues about the missing baryon problem \cite{2003ApJ...598L..79I,2004MNRAS.348..999I,2014ApJ...780L..33M,2015aska.confE..55M,2019ApJ...872...88R}. As FRBs may originate at high redshifts, they may also constitute a tool to obtain constraints on cosmological parameters, such as the equation of state of dark energy \cite{2014PhRvD..89j7303Z,2019PhRvD.100h3533K,2019PhRvD..99l3517L,2020ApJ...903...83Z,2022JCAP...02..006Q,2023JCAP...04..022Z}, the baryon density \cite{2014ApJ...788..189G,2016ApJ...830L..31Y,2018ApJ...856...65W,2019MNRAS.484.1637J,2023SCPMA..6620412Z}, or the Hubble parameter \cite{2022MNRAS.516.4862J,2022MNRAS.515L...1W,2022MNRAS.511..662H,2022arXiv221213433Z,2023ApJ...946L..49L,2024MNRAS.527.7861G}. FRB data may also be used to test the equivalence principle \cite{2015PhRvL.115z1101W,2016ApJ...821L...2N,2016ApJ...820L..31T,2022MNRAS.512..285R}, and primordial non-Gaussianity \cite{2021PhRvD.103b3517R}.

  As a cosmological distance measure in the absence of redshift information, DM is called \textit{standard ping}~\cite{2015PhRvL.115l1301M,2020PhRvD.102b3528R}, and enables us to construct three-dimensional maps of FRB catalogs. However, the DM receives a significant stochastic contribution from the propagation of the FRB pulse through the inhomogeneous universe, and is therefore far from a perfect distance proxy. The corresponding DM-based 3D map of the large-scale structure therefore appears to be distorted in the radial direction, the so-called \textit{dispersion measure space distortions}. The primary contributor to the distortions is the inhomogeneous distribution of free electrons~\cite{2014ApJ...780L..33M,2015PhRvL.115l1301M,2019PhRvD.100h3533K,2020PhRvD.102b3528R}. On top of this contribution, Ref.~\cite{2021PhRvD.103l3544A} has shown that special and general relativistic effects also induce additional distortions (see also Ref.~~\cite{2024arXiv240108862X} for the Shapiro time delay effect of dark matter substructure on the variance of the DM). Restricting ourselves to two-dimensional statistics, such as the angular power spectrum of projected DM observations, the relativistic effects remain a minor contributor that is likely unobservable.

  The impact of relativistic effects and redshift-space distortions (RSDs) in galaxy redshift surveys has been investigated~\cite{1987MNRAS.228..653S,2000ApJ...537L..77M,2011PhRvD..84d3516C,2011PhRvD..84f3505B}. The characteristic signal of relativistic effects in RSDs is the asymmetric distortions along the line-of-sight direction~\cite{2009JCAP...11..026M,2013MNRAS.434.3008C,2012arXiv1206.5809Y,2018JCAP...03..019T,2014PhRvD..89h3535B}, suggesting that investigating the galaxy clustering in the three-dimensional space is crucial to target relativistic effects.
  
  Motivated by the above, in this paper we investigate the anisotropy of the three-dimensional clustering in DM space, including both the inhomogeneous free electron distribution and relativistic effects based on Refs.~\cite{2015PhRvL.115l1301M,2020PhRvD.102b3528R,2021PhRvD.103l3544A}. Analogous to RSDs, relativistic effects produce an asymmetric clustering with respect to the line-of-sight direction. This asymmetry is characterized by a dipole in the multipole expansion of the correlation function. In this paper we investigate the behavior of the dipole anisotropy as well as the even multipole anisotropy. The former captures the impact of the relativistic effects, while the latter captures the impact of the inhomogeneous distribution of free electrons, both of which are complementary and therefore important observables for testing cosmological models and gravity theories. Based on the derived analytical expression for the multipoles, we also discuss the future detectability in an SKA-like survey for the multipoles of the FRBs cross-power spectra and in the galaxy-FRB cross-power spectra. Our investigation mainly focuses on the FRBs but applies to other standard pings (i.e., other bright radio transients having short time scales).

  This paper is organized as follows. In Sec.~\ref{sec: dispersion measure space}, we briefly introduce DM-based observations in a perturbed Friedmann-Lema\^{i}tre-Robertson-Walker (FLRW) universe, and present the analytical expression of the density field in DM space, following Ref.~\cite{2021PhRvD.103l3544A}. In Sec.~\ref{sec: FRB xi} and Sec.~\ref{sec: FRBxgal xi}, we consider the cross-correlation between FRBs with different biases and the cross-correlation between galaxies and FRBs, respectively. In these sections, we present the analytical expression of the correlation function in DM space, and numerically investigate their behavior. In Sec.~\ref{sec: future}, we forecast the expected signal-to-noise ratio for these measurements assuming SKA-like survey specifications. We conclude and summarize our results in Section~\ref{sec: summary}. We adopt the distant-observer limit throughout the analysis but discuss the wide-angle correction in Appendix~\ref{sec: app wa}. We also investigate the correction from the non-linear matter power spectrum to the dipole signal in Appendix~\ref{sec: NL dipole}. Throughout this paper, we apply the Einstein summation convention for repeated Greek and Latin indices, running from 0 to 3 and from 1 to 3, respectively. We work in units $c=\hbar=1$.

%%%%%%%%%%%%%%%%%%%%%%%%%%%%%%%%%%%%%%%%%%
%%%%%%%%%%%%%%%%%%%%%%%%%%%%%%%%%%%%%%%%%%
%%%%%%%%%%%%%%%%%%%%%%%%%%%%%%%%%%%%%%%%%%
\section{Three-dimensional clustering in DM space\label{sec: dispersion measure space}}
%%%%%%%%%%%%%%%%%%%%%%%%%%%%%%%%%%%%%%%%%%
%%%%%%%%%%%%%%%%%%%%%%%%%%%%%%%%%%%%%%%%%%
%%%%%%%%%%%%%%%%%%%%%%%%%%%%%%%%%%%%%%%%%%

  As the DM of electromagnetic pulses depends on the cumulative amount of plasma along the trajectory, we can regard it as a radial distance proxy and reconstruct the three-dimensional clustering of pulse emitters. However, the DM is not a perfect proxy for a radial distance because of inhomogeneities in the electron density~(e.g., Refs.~\cite{2014ApJ...780L..33M,2015PhRvL.115l1301M,2019PhRvD.100h3533K,2020PhRvD.102b3528R}). In addition to the electron density inhomogeneities, special and general relativistic effects have an impact on the trajectories and energies of the emitted photons, and thus the DM is further modulated. Hence the observed DM-based clustering pattern appears to be distorted~\cite{2021PhRvD.103l3544A}. In this section, we briefly review the observed density field in DM space following in the footsteps of Ref.~\cite{2021PhRvD.103l3544A}. Throughout our analysis, we simply assume that DM is directly related to the number density of cold electron plasma in the intergalactic medium, and ignore the local contributions (i.e. free electrons of host galaxies and in the Milky Way). This systematic would introduce suppression of the signal at small radial scales if the local contribution does not correlate to the cosmological signal in the intergalactic medium~\cite{2021JCAP...01..036N}.

  The DM of pulses in a Lorentz invariant form is given by~\cite{2021PhRvD.103l3544A}
  \begin{align}
    \mathcal{D} = \int^{\lambda_{\rm o}}_{\lambda_{\rm e}}{\rm d}\lambda\,
    \left.\left( n_{\rm e}\frac{k^{\mu}u_{{\rm e},\mu}}{1+z} \right)\right|_{\lambda} . \label{eq: def DM}
  \end{align}
  Here we define the redshift $z$, the 4-momentum of the massless pulse $k^{\mu} = {\rm d}x^{\mu}/{\rm d}\lambda$, with $\lambda$ being an affine parameter, the affine parameter at the observer $\lambda_{\rm o}$, the affine parameter at the source $\lambda_{\rm e}$, the 4-velocity of the source $u_{{\rm e},\mu}$, and the electron number density $n_{\rm e}$. The subscript $\lambda$ indicates that the integrand is evaluated along the trajectory of the pulse, corresponding to a null geodesic.

  Consider the perturbed FLRW metric:
  \begin{align}
    {\rm d}s^{2} = a^{2}\Bigl[ (1+2\psi){\rm d}\eta^{2} - (1-2\phi){\rm d}x^{2} \Bigr] ,
  \end{align}
  where the quantities $a$ and $\eta$ are the scale factor and the conformal time, respectively. Here, we consider only the scalar perturbations, $\psi$ and $\phi$.
  Up to the linear order in the perturbed variables, Eq.~(\ref{eq: def DM}) is given by~\cite{2021PhRvD.103l3544A}
  \begin{align}
    \mathcal{D}(\eta, \hat{\bm{n}}) &= \int^{\eta_{0}}_{\eta}{\rm d}\eta'\, a^{2}(\eta')\bar{n}_{\rm e}(\eta')
    \notag \\
    & \times 
    \Biggl[ 1 + \delta_{\rm e}(\eta', \chi' \hat{\bm{n}}) + 2\psi(\eta', \chi' \hat{\bm{n}})
    \notag \\
    & 
    + \int^{\eta_{0}}_{\eta'}{\rm d}\eta'' \left( \dot{\psi}(\eta'', \chi'' \hat{\bm{n}}) + \dot{\phi}(\eta'', \chi'' \hat{\bm{n}}) \right) \Biggr] , \label{eq: pertrubed DM}
  \end{align}
  where the quantities $\eta_{0}$ and $\hat{\bm{n}}$ are, respectively, the conformal time at the present and the unit vector $\hat{\bm{n}} = \bm{x}/|\bm{x}| = \bm{x}/\chi$ with $\chi = \eta_{0} -\eta$ being the radial comoving distance. A dot denotes a derivative with respect to the conformal time. Here, we decomposed the electron density into background and perturbation $n_{\rm e} = \bar{n}_{\rm e} (1+\delta_{\rm e})$. In deriving Eq.~(\ref{eq: pertrubed DM}), we solved the geodesic equation in the perturbed FLRW metric (see e.g., Refs.~\cite{2011PhRvD..84d3516C,2011PhRvD..84f3505B} and also Appendix~A in Ref.~\cite{2021PhRvD.103l3544A} in details). We note that ignoring the perturbations in Eq.~(\ref{eq: pertrubed DM}), we recover the background expression of the DM:
  \begin{align}
    \bar{\mathcal{D}}(z) &= \int^{z}_{0}{\rm d}z'\, \frac{(1+z')}{H(z')}a^{3}(z')\bar{n}_{\rm e}'(z') . \label{eq: DM bg}
  \end{align}

  Ref.~\cite{2021PhRvD.103l3544A} derived an analytical expression for the number density of observed sources in an observer's solid angle ${\rm d}\Omega_{o}$ and DM interval ${\rm d}\mathcal{D}$, in a manner analogous to the galaxy redshift survey~\cite{2011PhRvD..84d3516C,2011PhRvD..84f3505B}. We refer the readers to Ref.~\cite{2021PhRvD.103l3544A} for the details of the derivation and present only the final result. Assuming $\phi = \psi$, the expression of the fluctuation of the number density in DM space $\delta^{(\mathcal{D})}_{\rm X}$ for the species X is given by
%===========
\begin{align}
\delta^{(\mathcal{D})}_{\rm X} &= \delta_{\rm X} - \delta_{\rm e}
+ \bm{v}\cdot \hat{\bm{n}}
- 2 \hat{\nabla}^{2}\int^{\chi}_{0}{\rm d}\chi'\, \frac{\chi - \chi'}{\chi\chi'}\phi
\notag \\
& 
+ \mathcal{A}_{\rm X}(\chi)
\frac{\mathcal{H}}{a^{2}\bar{n}_{\rm e}} \int^{\eta_{0}}_{\eta}{\rm d}\eta'\, a^{2}\bar{n}_{\rm e}
\left( \delta_{\rm e} + 2\phi + 2\int^{\eta_{0}}_{\eta'}{\rm d}\eta''\, \dot{\phi}
\right)
\notag \\
& 
 - 3\phi - 2\int^{\eta_{0}}_{\eta}{\rm d}\eta'\,  \dot{\phi}
+ \frac{4}{\chi}\int^{\eta_{0}}_{\eta}{\rm d}\eta'\, \phi
,
\label{eq: DM space delta}
\end{align}
%===========
where the quantities $\delta_{\rm X}$, $\bm{v}$, and $\mathcal{H} = \dot{a}/a$ are the density fluctuation of the source number density for the species X, the source velocity, and the reduced Hubble parameter, respectively.
The operators $\hat{\bm{\nabla}}$ and $\hat{\nabla}^{2}$ are the angular gradient and angular Laplacian, respectively. These operators satisfy the following relations: $\hat{\nabla}_{i}\hat{n}_{j} = -\hat{n}_{i}\hat{n}_{j} + \delta_{ij}$ and $\hat{\nabla}^{2} \hat{\bm{n}} = - 2\hat{\bm{n}}$.
The quantity $\mathcal{A}_{\rm X}(\chi)$ defined in Eq.~(\ref{eq: DM space delta}) is explicitly given by
%===========
\begin{align}
\mathcal{A}_{\rm X}(\chi) = 1+f_{\rm X}-f_{\rm e}-\frac{2}{\mathcal{H}\chi} , \label{eq: cal A}
\end{align}
%===========
with $f_{i}\equiv {\rm d}\ln{(a^{3}\bar{n}_{i})}/{\rm d}\ln{a}$ being the evolution bias for the species $i$.

Equation~(\ref{eq: DM space delta}) is our starting expression of the DM space density. The third and fourth terms on the right-hand side are, respectively, a velocity term arising from the distortions of the observed comoving volume and weak lensing term. Here we simply ignored the magnification bias, which changes the prefactor of the weak lensing term. The second and third lines, respectively, correspond to the non-local density term arising from the light-cone integral of the DM and the other relativistic terms including the Sachs-Wolfe, Shapiro time delay, and integrated Sachs-Wolfe effects.

Importantly, the contributions proportional to $\mathcal{O}\left(\left(\mathcal{H}/k\right)^{2}\right)$ are suppressed with respect to the density and weak lensing terms. Ignoring these subdominant contributions, Eq.~(\ref{eq: DM space delta}) becomes
%===========
\begin{align}
\delta^{(\mathcal{D})}_{\rm X}(\eta,\bm{x}) &=
\delta^{\rm l}_{\rm X}(\eta,\bm{x})
+ \delta^{\rm v}_{\rm X}(\eta,\bm{x})
+ \delta^{\rm nl}_{\rm X}(\eta,\bm{x})
+ \delta^{\kappa}_{\rm X}(\eta,\bm{x}) , \label{eq: dens D}
\end{align}
%===========
where we define 
%===========
\begin{align}
\delta^{\rm l}_{\rm X}(\eta,\bm{x}) & = \left( b_{\rm X}-b_{\rm e} \right)\delta_{\rm L}(\eta,\bm{x}) , \label{eq: local}\\
\delta^{\rm v}_{\rm X}(\eta,\bm{x}) & = \bm{v}(\eta,\bm{x})\cdot \hat{\bm{n}} ,
\label{eq: delta v} \\
\delta^{\rm nl}_{\rm X}(\eta,\bm{x}) & = \mathcal{A}_{\rm X}(\chi)b_{\rm e} \frac{\mathcal{H}}{a^{2}\bar{n}_{\rm e}} \int^{\eta_{0}}_{\eta}{\rm d}\eta'\, a^{2}\bar{n}_{\rm e}
\delta_{\rm L} (\eta',\chi'\hat{\bm{n}}) , \label{eq: nl}\\
\delta^{\kappa}_{\rm X}(\eta,\bm{x}) & = - 2 \hat{\nabla}^{2}\int^{\chi}_{0}{\rm d}\chi'\, \frac{\chi - \chi'}{\chi\chi'}\phi(\eta',\chi'\hat{\bm{n}})
.
\label{eq: kappa}
\end{align}
%===========
We assume a linear bias relation for the source and electron densities, in which each density field is related to the linear density field $\delta_{\rm L}$ via $\delta_{\rm X} = b_{\rm X} \delta_{\rm L}$ and $\delta_{\rm e} = b_{\rm e} \delta_{\rm L}$.
The first to fourth terms on the right-hand side in Eq.~(\ref{eq: dens D}) correspond to the local density, local velocity, non-local density, and lensing contributions, respectively.
We evaluate the integrands of the non-local terms in Eq.~(\ref{eq: nl}) and (\ref{eq: kappa}) along the null geodesic, with $\chi' = \eta_{0} - \eta'$.

Moving into Fourier space, Eqs.~(\ref{eq: local})--(\ref{eq: kappa}) become
%===========
\begin{align}
\delta^{\rm l}_{\rm X}(\eta, \bm{x}) & = \left( b_{\rm X}-b_{\rm e} \right)
\int\frac{{\rm d}^{3}\bm{k}}{(2\pi)^{3}}e^{i\bm{k}\cdot\bm{x}}
\delta_{\rm L}(\eta, \bm{k}) , \label{eq: d l} \\
\delta^{\rm v}_{\rm X}(\eta, \bm{x}) & =
\int\frac{{\rm d}^{3}\bm{k}}{(2\pi)^{3}}e^{i\bm{k}\cdot\bm{x}}
\left( i \hat{\bm{k}}\cdot \hat{\bm{n}} \right)
\frac{\mathcal{H}f}{k} \delta_{\rm L}(\eta, \bm{k}) , \label{eq: d v}\\
\delta^{\rm nl}_{\rm X}(\eta, \bm{x}) & = \mathcal{A}_{\rm X}(\chi)b_{\rm e} \frac{\mathcal{H}}{a^{2}\bar{n}_{\rm e}} \int^{\chi}_{0}{\rm d}\chi'\, a^{2}(\eta')\bar{n}_{\rm e}(\eta')
\notag \\
& \times
\int\frac{{\rm d}^{3}\bm{k}}{(2\pi)^{3}}e^{i\bm{k}\cdot\left( \chi' \hat{\bm{n}} \right)}\delta_{\rm L} (\eta', \bm{k}) ,
\label{eq: d nl}
\\
\delta^{\kappa}_{\rm X}(\eta, \bm{x})
& = 
3\Omega_{\rm m0}\mathcal{H}^{2}_{0}
\int^{\chi}_{0}{\rm d}\chi'\, \frac{\chi - \chi'}{\chi\chi'}
\notag \\
& \times
\int\frac{{\rm d}^{3}\bm{k}}{(2\pi)^{3}}e^{i\bm{k}\cdot\left( \chi' \hat{\bm{n}} \right)}
\frac{-\chi'^{2}k^{2}_{\perp} - 2i\chi' k_{\parallel}}{a(\eta')k^{2}}
\delta_{\rm L} (\eta', \bm{k}) .
\label{eq: d kappa}
\end{align}
%===========
where the quantities $\Omega_{\rm m0}$, $\mathcal{H}_{0}$, and $f$ are the matter density parameter, reduced Hubble parameter at the present time, and the linear growth rate. $f$ is defined by $f\equiv {\rm d}\ln{D_{+}}/{\rm d}\ln{a}$ with $D_{+}$ the linear growth factor. We assume self-similar growth in the linear regime, such that $\delta_{\rm L}(\eta, \bm{k}) = D_{+}(\eta)\delta_{\rm L}(\eta_{0}, \bm{k})$.
We define the longitudinal and transverse components of the wave vector as $k_{\parallel} = \bm{k}\cdot\bm{\hat{n}}$ and $k_{\perp, i} = \left( \delta_{ia} - \hat{n}_{i}\hat{n}_{a}\right)k_{a}$, respectively.

In deriving the expressions above, we assume the following transfer functions:
%===========
\begin{align}
  \phi(\eta,\bm{k}) & = -\frac{3\Omega_{\rm m0}\mathcal{H}^{2}_{0}}{2ak^{2}}  \delta_{\rm L}(\eta, \bm{k}) , \\
  \bm{v}(\eta, \bm{k})& = i\hat{\bm{k}}\, \frac{\mathcal{H}f}{k} \delta_{\rm L}(\eta, \bm{k}) .
\end{align}
%===========
Equation~(\ref{eq: dens D}) together with Eqs.~(\ref{eq: d l})--(\ref{eq: d kappa}) presents the relation between the observed density fluctuation in DM space and real space density field. These expressions are key ingredients to compute the two-point statistics of the clustering of FRBs in DM space.

%%%%%%%%%%%%%%%%%%%%%%%%%%%%%%%%%%%%%%%%%%
%%%%%%%%%%%%%%%%%%%%%%%%%%%%%%%%%%%%%%%%%%
%%%%%%%%%%%%%%%%%%%%%%%%%%%%%%%%%%%%%%%%%%
\section{FRB correlation function\label{sec: FRB xi}}
%%%%%%%%%%%%%%%%%%%%%%%%%%%%%%%%%%%%%%%%%%
%%%%%%%%%%%%%%%%%%%%%%%%%%%%%%%%%%%%%%%%%%
%%%%%%%%%%%%%%%%%%%%%%%%%%%%%%%%%%%%%%%%%%

In this section, starting from the expression for the density field in DM space presented above, we investigate the FRB correlation function in DM space.
We consider the general situation where we can observe two populations of FRBs, X and Y, and we study the correlation of their overdensities measured at DM-space positions $\bm{x}_{1}$ and $\bm{x}_{2}$, respectively (which we denote $\delta^{(\mathcal{D})}_{\rm X}(\bm{x}_{1})$ and $\delta^{(\mathcal{D})}_{\rm Y}(\bm{x}_{2})$). Separating FRBs into different populations is possible, for example, if their host galaxies can be identified and split by brightness or color, leading to populations with different effective biases~(e.g., Ref.~\cite{2017MNRAS.470.2822A}). The exact procedure to separate these two populations with sufficiently different biases will be subject to statistic and systematic uncertainties. Here we will simply assume that such a split is possible and make predictions for the resulting cross-correlation function. Later on we will present the cross-correlation between galaxies and FRBs, which would not be significantly affected by these uncertainties.

We compute the cross-correlation:
%===========
\begin{align}
\xi_{\rm XY}(\bm{x}_{1},\bm{x}_{2}) & = 
\Braket{\delta^{(\mathcal{D})}_{\rm X}
(\bm{x}_{1})
\delta^{(\mathcal{D})}_{\rm Y}
(\bm{x}_{2})} , \label{eq: def xi}
\end{align}
%===========
with $\Braket{\cdots}$ denoting the ensemble average. We omit all time dependence for brevity. In computing the correlation function, we assume the distant-observer limit, i.e., $\hat{\bm{n}}_{1} = \hat{\bm{n}}_{2} = \hat{\bm{z}}$ with $\hat{\bm{z}}$ being a fixed line-of-sight vector. Under the distant-observer limit, the correlation function is given as a function of the separation $s = |\bm{s}| = |\bm{x}_{2} - \bm{x}_{1}|$ and the directional cosine $\mu$ between $\hat{\bm{z}}$ and $\hat{\bm{s}}$. For the non-local terms, we use the Limber approximation~\cite{1953ApJ...117..134L} (see Refs.~\cite{2000ApJ...537L..77M,2004PhRvD..69h3524A,2007PhRvD..76j3502H,2018JCAP...03..019T}). In Appendix~\ref{sec: app wa}, we discuss the wide-angle correction beyond the distant-observer limit, which is, however, negligibly small.

Substituting Eq.~(\ref{eq: dens D}) into Eq.~(\ref{eq: def xi}), we derive the analytical expression for the correlation function. We split the nonvanishing contributions into eight pieces:
%===========
\begin{align}
\xi_{\rm XY} &=
\left(
\xi^{{\rm l}\times{\rm l}}_{\rm XY} + \xi^{{\rm l}\times{\rm v}}_{\rm XY} + \xi^{{\rm v}\times{\rm v}}_{\rm XY}  \right)
\notag \\
& 
+ \left( \xi^{{\rm nl}\times{\rm nl}}_{\rm XY} + \xi^{{\rm nl}\times{\kappa}}_{\rm XY} + \xi^{{\kappa}\times{\kappa}}_{\rm XY} \right) 
+ \left( \xi^{{\rm l}\times{\rm nl}}_{\rm XY} + \xi^{{\rm l}\times{\kappa}}_{\rm XY} \right)
, \label{eq: xi FRB}
\end{align}
%===========
where we define 
%===========
\begin{align}
\xi^{{\rm a}\times {\rm b}}_{\rm XY} & = 
\begin{cases}
\Braket{\delta^{\rm a}_{\rm X}
(\bm{x}_{1})
\delta^{\rm b}_{\rm Y}
(\bm{x}_{2})}
& ({\rm a} = {\rm b})
\\
\Braket{\delta^{\rm a}_{\rm X}
(\bm{x}_{1})
\delta^{\rm b}_{\rm Y}
(\bm{x}_{2})}
+ ({\rm a}\leftrightarrow {\rm b})
& ({\rm a} \neq {\rm b})
\end{cases}
. 
\end{align}
%===========
The first, second, and third parentheses in the right-hand side of Eq.~(\ref{eq: xi FRB}) stand for the pure local term arising from the local density and local velocity terms, the pure non-local term arising from the non-local density and weak lensing terms, and the cross-correlation between local and non-local terms, respectively. 
The cross-correlations between the local velocity and non-local terms vanish, i.e., $\Braket{\delta^{\rm v}_{\rm X}\delta^{\rm nl}_{\rm Y}} = \Braket{\delta^{\rm v}_{\rm X}\delta^{\rm \kappa}_{\rm Y}} = 0$, because the parallel components to the line-of-sight direction do not contribute to the correlation function in the flat-sky, Limber approximation~(see, e.g., Refs.~\cite{2018JCAP...03..019T,2018JCAP...10..032T,2021JCAP...01..036N,2021JCAP...07..045J}).

The explicit forms of the right-hand side of Eq.~(\ref{eq: xi FRB}) are given by
%===========
\begin{align}
\xi^{{\rm l}\times {\rm l}}_{\rm XY}
& = 
(b_{\rm X}-b_{\rm e})(b_{\rm Y}-b_{\rm e})\Xi^{(0)}_{0}(\eta,s)
, 
\label{eq: l-l}\\
%%%%%%%%%%%%%%%%%%%%
%%%%%%%%%%%%%%%%%%%%
\xi^{{\rm l}\times {\rm v}}_{\rm XY}
& = 
- (b_{\rm X} - b_{\rm Y})
\mu \left( s\mathcal{H}f \right) \Xi^{(1)}_{1}(\eta,s)
, \label{eq: l-v}\\
%%%%%%%%%%%%%%%%%%%%
%%%%%%%%%%%%%%%%%%%%
\xi^{{v}\times {v}}_{\rm XY}
& = 
(s\mathcal{H}f)^{2} \left[ 
\left( \frac{1}{3}-\mu^{2} \right) \Xi^{(2)}_{2}(\eta,s)
+ \frac{1}{3}\Xi^{(2)}_{0}(\eta,s)\right]
,
\label{eq: v-v}
\end{align}
%===========
for the purely local contributions, 
%===========
\begin{align}
\xi^{{\rm nl}\times {\rm nl}}_{\rm XY}
& = 
\mathcal{A}_{\rm X}(\chi)\mathcal{A}_{\rm Y}(\chi)\left( b_{\rm e}\frac{\mathcal{H}}{a^{2}\bar{n}_{\rm e}} \right)^{2}
\int^{\chi}_{0}{\rm d}\chi'\, 
\notag \\
&
 \times 
\left( a^{2}(\eta')\bar{n}_{\rm e}(\eta') \right)^{2}
\, 
\mathcal{J}\left(\eta', \frac{\chi'}{\chi}s\sqrt{1-\mu^{2}} \right)
, \label{eq: nl-nl 1} \\
%%%%%%%%%%%%%%%%%%%%
%%%%%%%%%%%%%%%%%%%%
\xi^{{\rm nl}\times {\kappa}}_{\rm XY}
& = 
- \left( \mathcal{A}_{\rm X}(\chi) + \mathcal{A}_{\rm Y}(\chi) \right)b_{\rm e}\frac{\mathcal{H}}{a^{2}\bar{n}_{\rm e}}
\frac{3\Omega_{\rm m0}\mathcal{H}^{2}_{0}}{\chi}
\int^{\chi}_{0}{\rm d}\chi'\,
\notag \\
&
\times 
a(\eta')\bar{n}_{\rm e}(\eta')
\left( \chi - \chi' \right) \chi'
\, 
\mathcal{J}\left( \eta', \frac{\chi'}{\chi}s\sqrt{1-\mu^{2}} \right)
, \\
%%%%%%%%%%%%%%%%%%%%
%%%%%%%%%%%%%%%%%%%%
\xi^{\kappa \times \kappa}_{\rm XY}
& = 
\left( \frac{3\Omega_{\rm m0}\mathcal{H}^{2}_{0}}{\chi} \right)^{2}
\int^{\chi}_{0}{\rm d}\chi'\,
\notag \\
& \times 
\left(  \frac{(\chi - \chi') \chi'}{a(\eta')}\right)^{2}\, 
\mathcal{J}\left(\eta', \frac{\chi'}{\chi}s\sqrt{1-\mu^{2}} \right)
,
\label{eq: nl-nl 2}
\end{align}
%===========
for the purely non-local contributions, and
%===========
\begin{align}
\xi^{{\rm l}\times {\rm nl}}_{\rm XY}
& = 
\Bigl( 
\Theta(\chi_{2}-\chi_{1})(b_{\rm X}-b_{\rm e})
\mathcal{A}_{\rm Y}
\notag \\
& 
+ \Theta(\chi_{1}-\chi_{2})(b_{\rm Y}-b_{\rm e})
\mathcal{A}_{\rm X}
\Bigr)
b_{\rm e}\mathcal{H}
\mathcal{J}(\eta, s\sqrt{1-\mu^{2}})
, \label{eq: l-nl 1}\\
%%%%%%%%%%%%%%%%%%%%
%%%%%%%%%%%%%%%%%%%%
\xi^{{\rm l}\times {\kappa}}_{\rm XY}
& = 
\Bigl( 
- \Theta(\chi_{2}-\chi_{1})(b_{\rm X}-b_{\rm e})
+ \Theta(\chi_{1}-\chi_{2})(b_{\rm Y}-b_{\rm e})
\Bigr)
\notag \\
& 
\times 
\frac{3\Omega_{\rm m0}\mathcal{H}^{2}_{0}}{a}
s\mu
\mathcal{J}(\eta,s\sqrt{1-\mu^{2}})
\label{eq: l-kappa}
,
\end{align}
%===========
for the cross-correlation between them. In the above, we define the following functions:
%===========
\begin{align}
\Xi^{(n)}_{\ell}(\eta, s) &= \int\frac{k^{2}{\rm d}k}{2\pi^{2}}
\frac{j_{\ell}(ks)}{(ks)^{n}} P_{\rm L}(\eta,k) , \label{eq: def Xi} \\
\mathcal{J}(\eta, s) & = 
\int\frac{{\rm d}k}{2\pi} k J_{0}\left( k s \right)
P_{\rm L}(\eta, k) \label{eq: def calJ}
\end{align}
%===========
where the functions $j_{\ell}$ and $J_{0}$ are, respectively, the spherical Bessel function of order $\ell$, and the Bessel function of the first kind of order zero. We define the linear matter power spectrum at the time $\eta$: $\Braket{\delta_{\rm L}(\eta, \bm{k})\delta_{\rm L}(\eta, \bm{k}')} = (2\pi)^{3}\delta^{3}_{\rm D}(\bm{k}+\bm{k}') P_{\rm L}(\eta, k)$.

%=================================================
\begin{figure*}
\centering
\includegraphics[width=0.95\textwidth]{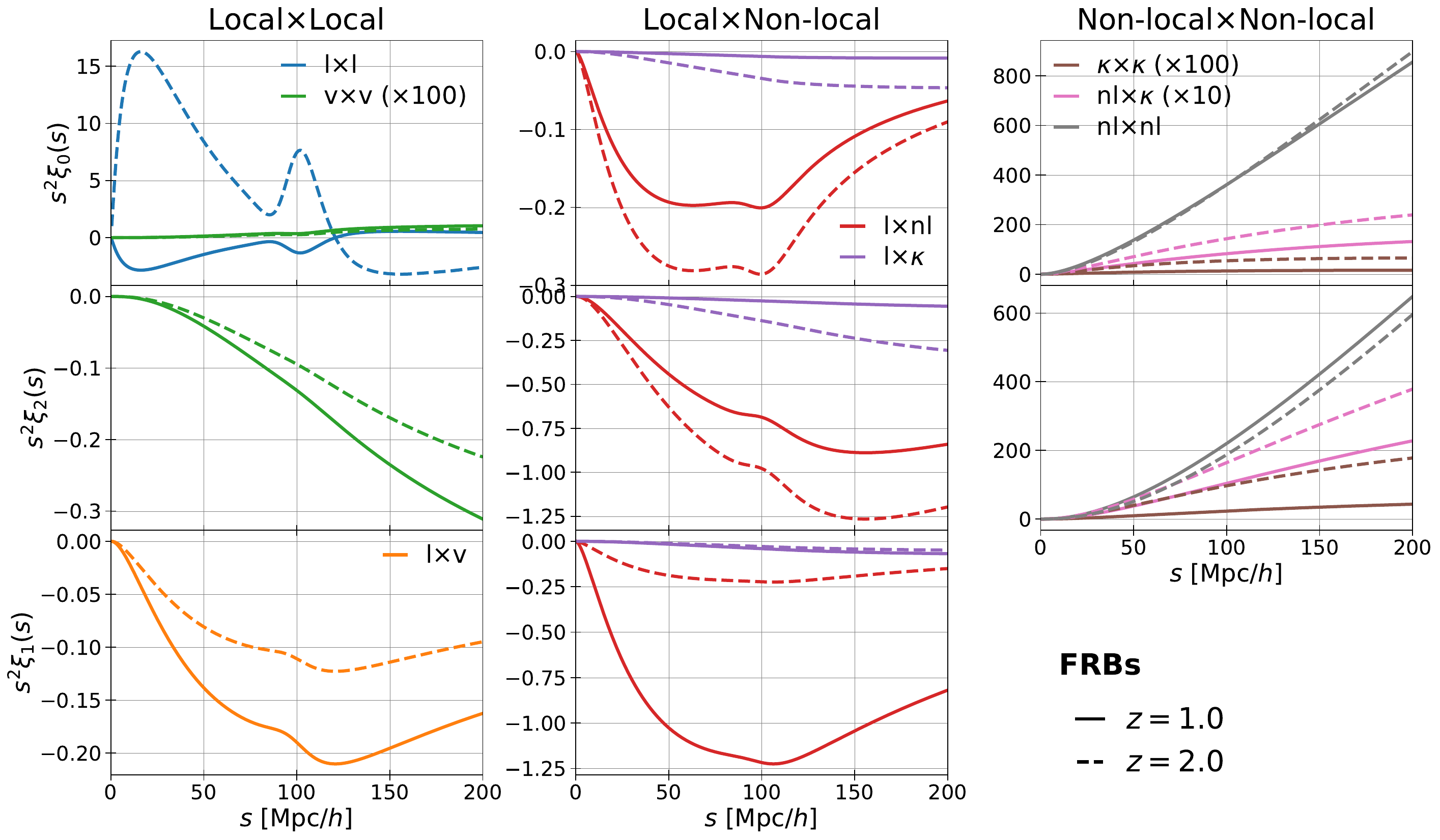}
\caption{Contributions to the multipoles at $z=1.0$ (solid lines) and $z=2.0$ (dashed lines). From left to right, we show the pure local, pure non-local, and crosstalk between the local and non-local contributions, respectively. From top to bottom, we present the monopole ($\ell=0$), quadrupole ($\ell=2$), and dipole ($\ell=1$), respectively. For the bias parameter, we use Eqs.~(\ref{eq: bias SKA2})--(\ref{eq: bias bY}), and have $(b_{\rm X}, b_{\rm Y}) = (1.71, 0.71)$ for $z=1.0$ and $(b_{\rm X}, b_{\rm Y}) = (3.15, 2.15)$ for $z=2.0$. We set $f_{\rm s} = f_{\rm e} = 0$ and $b_{e} = 1$. For presentation purposes, the results for $\xi^{{\rm v}\times{\rm v}}_{\rm XY}$, $\xi^{\kappa\times \kappa}_{\rm XY}$, and $\xi^{{\rm nl}\times{\kappa}}_{\rm XY}$ are multiplied by 100, 100, and 10, respectively.}
\label{fig: FRBs xi multipoles}
\end{figure*}
%=================================================

As a consequence of the DM-space distortions, the derived expressions depend on the directional cosine $\mu$. As done in the standard analysis of galaxy clustering in redshift space, we quantify this anisotroy by using a multipole expansion of the correlation function, weighting with the Legendre polynomials $\mathcal{L}_{\ell}(\mu)$ with $\mu = \hat{\bm{s}}\cdot\hat{\bm{z}}$:
%===========
\begin{equation}
\xi_{\ell}(s)
= \frac{2\ell+1}{2}\int^{1}_{-1}{\rm d}\mu\; \xi_{\rm XY}(s,\mu) \mathcal{L}_{\ell}(\mu) . \label{eq: multipoles}
\end{equation}
%===========

In the local contributions (\ref{eq: l-l})--(\ref{eq: v-v}), the velocity term induces the anisotropy in the correlation function. This induced anisotropy comes not from the standard Kaiser effect, seen in redshift space, but due to observed volume distortions (i.e. a pure relativistic effect). Interestingly, the cross-correlation between the local density and local velocity given in Eq.~(\ref{eq: l-v}) exhibits the asymmetric clustering along the line-of-sight direction, i.e., contributes to a dipole ($\ell=1$) anisotropy. This dipole contribution is non-zero only when we cross-correlate samples with different biases, due to the prefactor $b_{\rm X}-b_{\rm Y}$. This effect is also reproduced in the redshift-space clustering of galaxies (see, e.g., Refs.~\cite{2009JCAP...11..026M,2012arXiv1206.5809Y,2013MNRAS.434.3008C,2014PhRvD..89h3535B,2014CQGra..31w4001Y,2017JCAP...01..032G,2018JCAP...03..019T,2019MNRAS.483.2671B,2020JCAP...07..048B,2020MNRAS.498..981S,2022MNRAS.511.2732S,2023MNRAS.524.4472S}).
The cross-correlation between the local and non-local contributions (\ref{eq: l-nl 1})--(\ref{eq: l-kappa}) produces both even and odd multipoles, while the pure non-local contributions (\ref{eq: nl-nl 1})--(\ref{eq: nl-nl 2}) produce only even multipoles because of their symmetric dependence on $\mu$. From these analytical expressions, the local contributions~(\ref{eq: l-l})--(\ref{eq: v-v}) induce only anisotropies with $\ell\leq2$, whereas the non-local contributions (\ref{eq: nl-nl 1})--(\ref{eq: l-kappa}) contribute to arbitrarily high multipoles.

In Fig.~\ref{fig: FRBs xi multipoles}, we numerically demonstrate the contributions to the first three multipoles, i.e., the monopole ($\ell=0$), dipole ($\ell=1$), and quadrupole ($\ell=2$). In this plot, we model the background electron density as \cite{2021PhRvD.103l3544A}
%===========
\begin{align}
\bar{n}_{\rm e} = \frac{3H^{2}_{0}\Omega_{\rm b0}}{8\pi G m_{\rm p}} \frac{x_{\rm e}(z)(1+x_{\rm H}(z))}{2}a^{-3} , 
\label{eq: ne com}
\end{align}
%===========
where we use $x_{\rm H} = 0.75$ and $x_{\rm e} = 1$ for simplicity.
Through Eq.~(\ref{eq: DM bg}) together with Eqs.~(\ref{eq: ne com}), the background DM is a monotonic function of redshift $z$. Therefore, to facilitate the interpretation of our results, throughout the rest of this paper, we present results using redshift as a time indicator in the lightcone.

We assume a mean bias given by the specifications of SKA2 HI galaxies given in Ref.~\cite{2016ApJ...817...26B}:
%===========
\begin{align}
b_{\rm SKA}(z) &= c_{4} \exp{\left( c_{5} z\right)}, 
\label{eq: bias SKA2}
\end{align}
%===========
with $c_{4} = 0.554$ and $c_{5}=0.783$. We note that there is no evidence that HI galaxies are the preferential hosts of FRBs, and we only use this parametrisation in order to produce results assuming realistic galaxy bias values. We split the full FRB host population into two sub-samples with different biases 
%===========
\begin{align}
b_{\rm X}(z) & = b_{\rm SKA}(z) + \Delta b/2 , \label{eq: bias bX}\\
b_{\rm Y}(z) & = b_{\rm SKA}(z) - \Delta b/2 . \label{eq: bias bY}
\end{align}
%===========
We adopt an arbitrary bias difference $b_{\rm X} - b_{\rm Y} = \Delta b = 1$, noting again that our results, particularly in the case of the FRB-FRB correlations, depend on our ability to select such sub-samples.

As shown in Fig. \ref{fig: FRBs xi multipoles}, we find that the even multipoles (top two panels) are dominated by the non-local contributions associated with the line-of-sight integrated electron overdensity, $\xi^{{\rm nl}\times{\rm nl}}_{\rm XY}$, and display significant large-scale power. The relative amplitude of this contribution is consistent with the results of the angular power spectrum shown in Ref.~\cite{2021PhRvD.103l3544A}. The local density contribution ($\xi^{{\rm l}\times{\rm l}}_{\rm XY}$) is a sub-dominant but non-negligible contribution to the monopole, depending on scales and redshift. We note that the local density contribution becomes negative for $z=1$ because it is proportional to $(b_{\rm X}-b_{\rm e})(b_{\rm Y}-b_{\rm e})$, and $b_{\rm Y} < b_{\rm e} = 1$ at $z=1$. Turning to the dipole, all three contributions to the dipole have similar amplitudes. Among them, the cross-correlation between the local and non-local density terms ($\xi^{{\rm l}\times{\rm nl}}_{\rm XY}$) is a major contributor at $z=1$ but is suppressed at $z=2$. Accordingly, the local velocity contribution, $\xi^{{\rm l}\times{\rm v}}_{\rm XY}$, may dominate the dipole at high redshifts.

%=================================================
\begin{figure}
\centering
\includegraphics[width=0.49\textwidth]{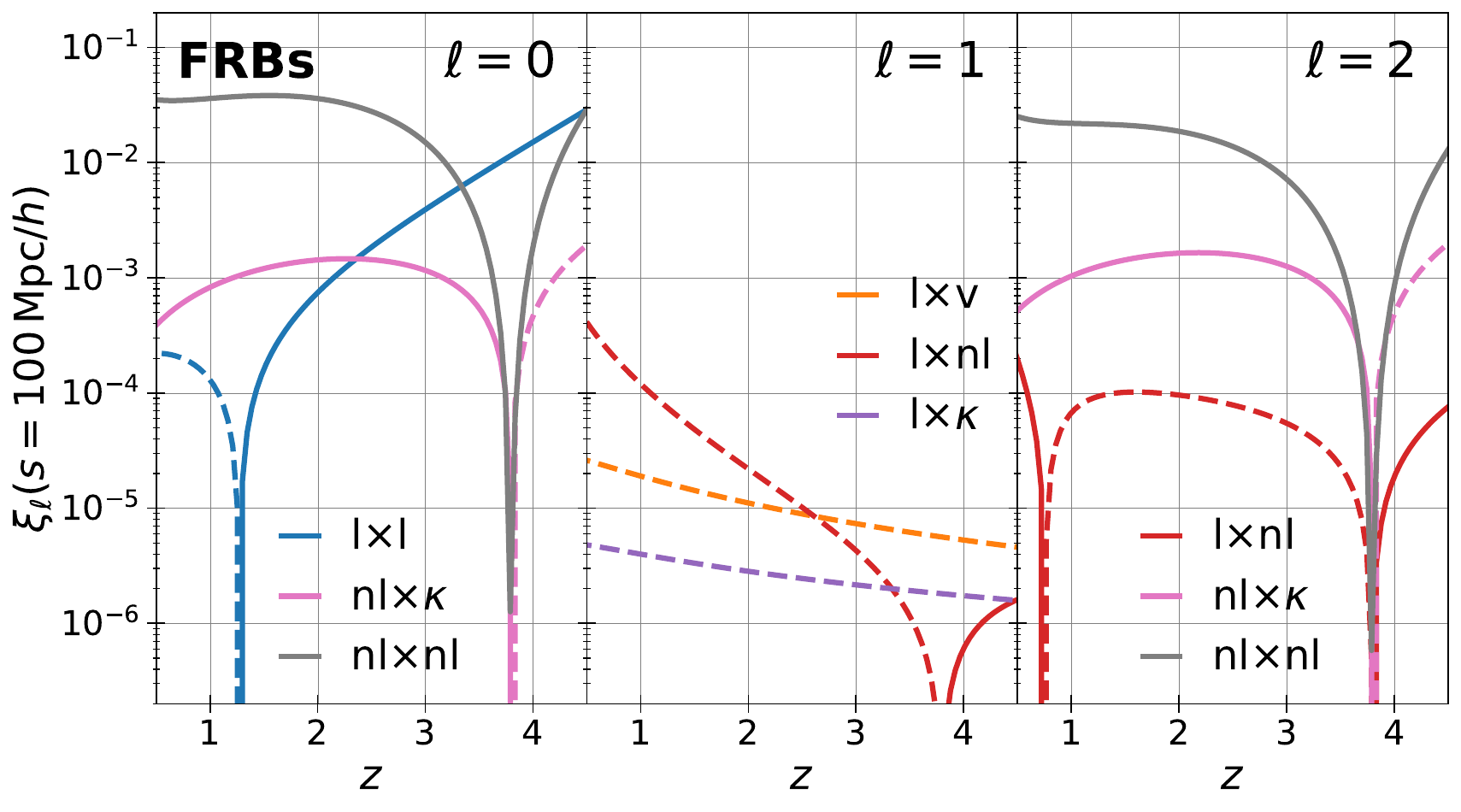}
\caption{First three dominant contributions to each multipole at $s=100\, {\rm Mpc}/h$ as a function of redshift. From left to right, we show the monopole, dipole, and quadrupole, respectively. The dashed lines indicate the negative amplitude.}
\label{fig: FRBs xi zred}
\end{figure}
%=================================================

Figure~\ref{fig: FRBs xi zred} shows the redshift dependence of the multipoles at $s = 100\, {\rm Mpc}/h$.
As we have already seen in Fig.~\ref{fig: FRBs xi multipoles}, the even multipoles at low redshift $z\lesssim 3$ are dominated by the non-local density term~($\xi^{{\rm nl}\times{\rm nl}}_{\rm XY}$). This makes it impossible to detect the BAO feature in the FRB correlation function, for example, which is only present in correlations involving local contributions. Around $z\approx 3.8$, the contributions including the non-local density term become zero because of the vanishing factor $\mathcal{A}_{\rm X/Y} = 0$ (see Eq.~\ref{eq: cal A}). 
Around this specific redshift, since the non-local density contribution vanishes at all scales, the local contributions start to dominate the multipoles.
We also find the other zero crossings at $z\approx 1.25$ in the monopole of $\xi^{{\rm l}\times{\rm l}}_{\rm XY}$ and $z\approx 0.75$ in the quadrupole of $\xi^{{\rm l}\times{\rm nl}}_{\rm XY}$, respectively.
These zero-crossings correspond to the redshift satisfying $b_{\rm Y}-b_{\rm e} = 0$ in Eq.~(\ref{eq: l-l}) and $(b_{\rm X}-b_{\rm e}) + (b_{\rm Y}-b_{\rm e}) = 0$ in Eq.~(\ref{eq: l-nl 1}) for the monopole and quadrupole, respectively. The local velocity contribution, which is always subdominant in the angular power spectrum~\cite{2021PhRvD.103l3544A}, dominates the dipole at $z\gtrsim 3$.
This would, in principle, allow us to isolate the local velocity term by studying three-dimensional clustering in DM space. It is important to emphasize that the occurrence of these zero-crossings, our ability to isolate the velocity dipole, and to detect the BAO at high redshifts, are entirely dependent on the parametrisations assumed for all the astrophysical quantities entering our prediction ($b_{\rm X}$, $b_{\rm e}$, $\bar{n}_{\rm e}$, $f_{\rm e}$, $f_{\rm X}$), and are therefore subject to the current large uncertainties in the properties of FRB hosts.

%=================================================
\begin{figure}
\centering
\includegraphics[width=0.49\textwidth]{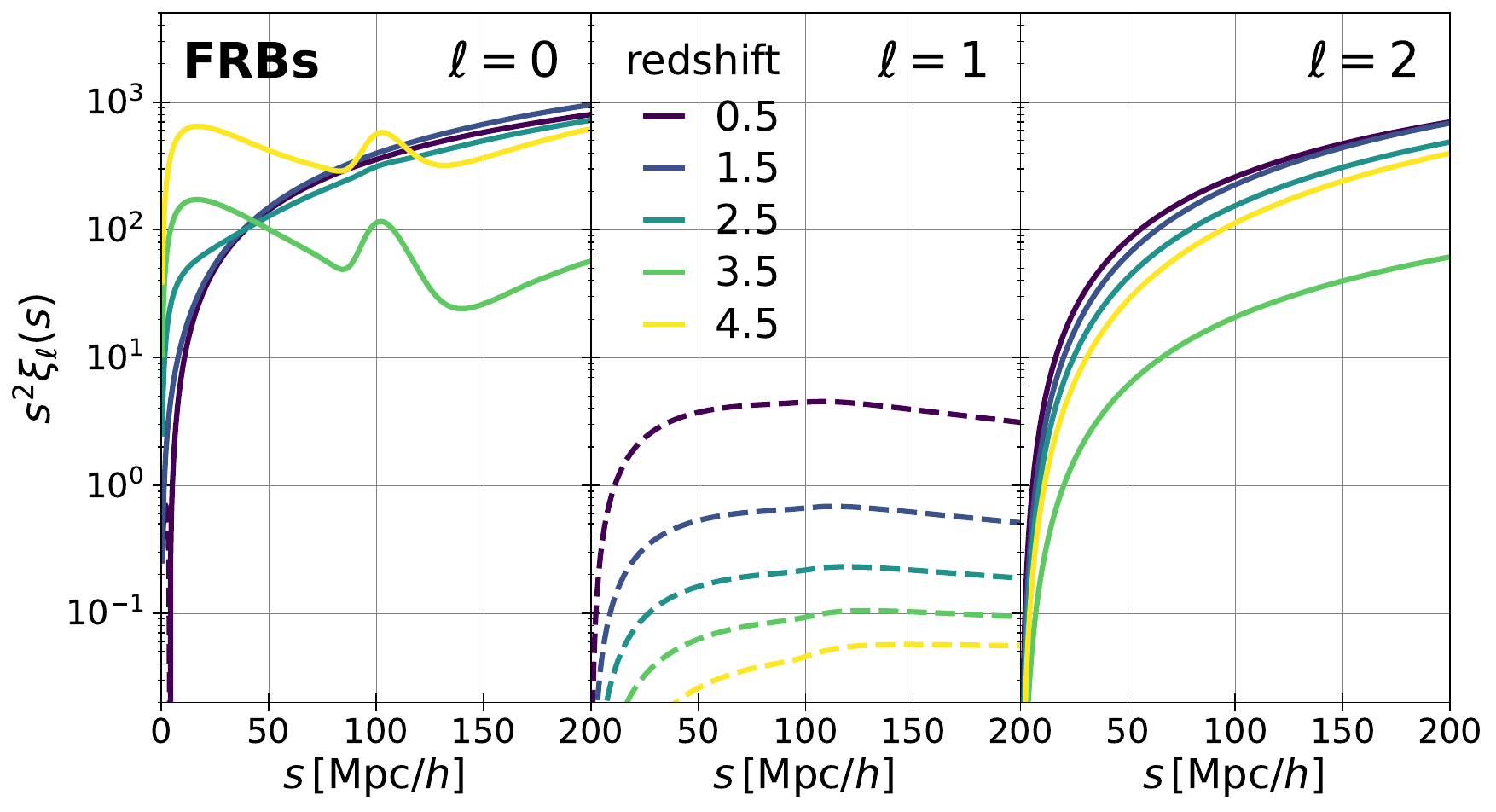}
\caption{Redshift evolution of the total signal of the multipoles. From left to right, we present the monopole, dipole, and quadrupole, respectively.  For the bias parameters, we use the same setup in Figs.~\ref{fig: FRBs xi multipoles} and \ref{fig: FRBs xi zred}. The dashed lines indicate the negative amplitude.}
\label{fig: FRBs xi multipoles total}
\end{figure}
%=================================================

Finally, we show the redshift dependence of the total signal for all multipoles in Fig.~\ref{fig: FRBs xi multipoles total}. The monopole shows a qualitatively different behavior at $z\lesssim 3.5$ and $z\gtrsim 3.5$, because the former is dominated by the non-local density contribution while the latter is dominated by the local density contribution, as seen in Fig.~\ref{fig: FRBs xi zred}. 
Turning to the dipole, it monotonically decreases with increasing redshifts, as again expected from Fig.~\ref{fig: FRBs xi zred}. The overall shape of the dipole does not drastically change because the two dominant contributions, $\xi^{{\rm l}\times{\rm v}}_{\rm XY}$ and $\xi^{{\rm l}\times{\rm nl}}_{\rm XY}$, display a similar behavior. As the quadrupole is mainly controlled by $\xi^{{\rm nl}\times{\rm nl}}_{\rm XY}$, it is suppressed around $z\approx 3.5$.

%%%%%%%%%%%%%%%%%%%%%%%%%%%%%%%%%%%%%%%%%%
%%%%%%%%%%%%%%%%%%%%%%%%%%%%%%%%%%%%%%%%%%
%%%%%%%%%%%%%%%%%%%%%%%%%%%%%%%%%%%%%%%%%%
\section{Galaxy-FRB cross-correlation function\label{sec: FRBxgal xi}}
%%%%%%%%%%%%%%%%%%%%%%%%%%%%%%%%%%%%%%%%%%
%%%%%%%%%%%%%%%%%%%%%%%%%%%%%%%%%%%%%%%%%%
%%%%%%%%%%%%%%%%%%%%%%%%%%%%%%%%%%%%%%%%%%

So far, we have investigated the anisotropy in the three-dimensional correlation function of FRBs measured in DM space, assuming that we obtain two different subsamples of FRBs, even though there is an inherent uncertainty in the splitting procedure. Moreover, the observed number of FRBs is expected to be smaller than that of galaxies. It is therefore interesting to consider the cross-correlation between FRBs and galaxies, which should be less sensitive to FRB shot noise and to our ability to sub-sample the FRB population in a meaningful way.

In the case of galaxies, the source's redshift is used as a distance proxy. The observed redshift is affected by various relativistic effects due to light propagation effects in an inhomogeneous universe, and these have been described and quantified in the literature (see, e.g., Refs.~\cite{1987MNRAS.228..653S,2004MNRAS.348..581P,2009PhRvD..80h3514Y,2000ApJ...537L..77M,2010PhRvD..82h3508Y,2011PhRvD..84f3505B,2011PhRvD..84d3516C,2014CQGra..31w4001Y,2018JCAP...05..043L,2020JCAP...07..048B}). 
Considering the dominant contributions at linear order, the observed galaxy overdensity field in redshift space is given by~\cite{2009PhRvD..80h3514Y,2011PhRvD..84d3516C,2011PhRvD..84f3505B}
%===========
\begin{align}
\delta^{({\rm S})}(\eta,\hat{\bm{n}}) & =
\delta^{\rm l}_{\rm g}(\eta,\hat{\bm{n}})
+ \delta^{\rm K}_{\rm g} (\eta,\hat{\bm{n}})
+ \delta^{\rm v}_{\rm g} (\eta,\hat{\bm{n}})
+ \delta^{\kappa}_{\rm g} (\eta,\hat{\bm{n}})
, \label{eq: RSD density}
\end{align}
%===========
where we define
%===========
\begin{align}
\delta^{\rm l}_{\rm g}(\eta,\hat{\bm{n}}) & = b_{\rm g}\delta_{\rm L}(\bm{x}) \label{eq: delta g}\\
\delta^{\rm v}_{\rm g} (\eta,\hat{\bm{n}}) &= \mathcal{B}(\chi) \delta^{\rm v} (\eta,\hat{\bm{n}}) , \\
\delta^{\rm K}_{\rm g} (\eta,\hat{\bm{n}}) & = - \frac{1}{\mathcal{H}} \left( \hat{\bm{n}}\cdot\bm{\nabla} \right)\left( \bm{v}\cdot{\hat{\bm{n}}} \right) \label{eq: delta Kaiser} , \\
\delta^{\kappa}_{\rm g} (\eta,\hat{\bm{n}}) & = \delta^{\kappa} (\eta,\hat{\bm{n}}) , \label{eq: delta kappa}
\end{align}
%===========
The quantities, $\delta^{\rm v}$ and $\delta^{\kappa}$, are defined in Eq.~(\ref{eq: delta v}) and (\ref{eq: kappa}), respectively. 
We introduced the linear galaxy bias parameter $b_{\rm g}$ and the time-dependent function $\mathcal{B}(\chi)$ given by
%===========
\begin{align}
\mathcal{B}(\chi) = 
\left( - 5s + \frac{5s-2}{\chi\mathcal{H}} - \frac{\dot{\mathcal{H}}}{\mathcal{H}^{2}} + f^{\rm ev}\right) ,
\end{align}
%===========
where the quantities $s$ and $f^{\rm ev}$ are the magnification bias and evolution bias of galaxies, respectively.
The terms on the right-hand side of Eq.~(\ref{eq: RSD density}) stand for the local density contribution, Kaiser effect~\cite{1987MNRAS.227....1K,1992ApJ...385L...5H}, and Doppler contributions, and the lensing magnification contribution, respectively.
Moving into Fourier space, the Kaiser term (\ref{eq: delta Kaiser}) becomes
%===========
\begin{align}
\delta^{\rm K}(\eta, \hat{\bm{n}}) & =
f \int\frac{{\rm d}^{3}k}{(2\pi)^{3}}e^{i\bm{k}\cdot\bm{x}} (\hat{\bm{k}}\cdot\hat{\bm{n}})^{2} \delta_{\rm L}(\bm{k},\eta) .
\end{align}
%===========
The expressions for $\delta^{\rm l}_{\rm g}$, $\delta^{\rm v}_{\rm g}$, and $\delta^{\kappa}_{\rm g}$ in Fourier space are given in Eqs.~(\ref{eq: d l}), (\ref{eq: d v}), and (\ref{eq: d kappa}), respectively.

Using this notation, we calculate the cross-correlation between galaxies and FRBs:
%===========
\begin{align}
\xi_{\rm gX}(\bm{x}_{1},\bm{x}_{2}) & = 
\Braket{\delta^{({\rm S})}_{\rm g}
(\bm{x}_{1})
\delta^{(\mathcal{D})}_{\rm X}
(\bm{x}_{2})} . \label{eq: xi g x FRB}
\end{align}
%===========
Here, in order to distinguish from the notation of the FRB cross-correlation function $\xi_{\rm XY}$, we denote the FRB-galaxy cross-correlation function $\xi_{\rm gX}$ where ${\rm X}$ represents FRBs having the bias parameter $b_{\rm X}$. We omit the time dependence in the notations for simplicity.

Substituting Eqs.~(\ref{eq: delta g})--(\ref{eq: delta kappa}) into Eq.~(\ref{eq: xi g x FRB}), we derive the analytical expression for the galaxy-FRB cross-correlation function. Similar to the FRB correlation function, we split the galaxy-FRB correlation function into 9 pieces:
%===========
\begin{align}
\xi_{\rm gX} =&
\left( \xi^{{\rm l}\times{\rm l}}_{\rm gX}
+ \xi^{{\rm l}\times{\rm v}}_{\rm gX}
+ \xi^{{\rm v}\times{\rm v}}_{\rm gX}
+ \xi^{{\rm K}\times{\rm l}}_{\rm gX}
+ \xi^{{\rm K}\times{\rm v}}_{\rm gX} \right)
\notag\\
&
+ \left( \xi^{{\rm nl}\times{\kappa}}_{\rm gX} + \xi^{{\kappa}\times{\kappa}}_{\rm gX} \right)
+ \left( \xi^{{\rm l}\times{\rm nl}}_{\rm gX} + \xi^{{\rm l}\times{\kappa}}_{\rm gX} \right)
, \label{eq: xi FRB x galaxy}
\end{align}
%===========
where we define 
%===========
\begin{align}
\xi^{{\rm a}\times {\rm b}}_{\rm gX} & = 
\Braket{\delta^{\rm a}_{\rm g}(\bm{x}_{1}) \delta^{\rm b}_{\rm X}(\bm{x}_{2})}
+ ({\rm a}\leftrightarrow {\rm b})
, 
\end{align}
%===========
for $({\rm a},{\rm b})=({\rm l},{\rm v}), ({\rm l},{\kappa}), ({\rm v},{\kappa})$, and otherwise
%===========
\begin{align}
\xi^{{\rm a}\times {\rm b}}_{\rm gX} & = 
\Braket{\delta^{\rm a}_{\rm g}(\bm{x}_{1})) \delta^{\rm b}_{\rm X}((\bm{x}_{2}))} .
\end{align}
%===========
In Eq.~(\ref{eq: xi FRB x galaxy}), the first, second, and third parentheses on the right-hand side, respectively, correspond to the pure local contributions, the pure non-local contributions, and the cross-correlation between local and non-local terms. Similar to the FRB cross-correlation function, the cross-contributions between the local velocity or Kaiser terms and non-local terms vanish in the Limber approximation, i.e., $\Braket{\delta^{\rm v}_{\rm g}\delta^{\rm nl}_{\rm X}} = \Braket{\delta^{\rm K}_{\rm g}\delta^{\rm nl}_{\rm X}} = \Braket{\delta^{\rm v}_{\rm g}\delta^{\kappa}_{\rm X}} = \Braket{\delta^{\rm K}_{\rm g}\delta^{\kappa}_{\rm X}} = 0$.
The explicit forms of each contribution are given by 
%===========
\begin{align}
\xi^{{\rm l}\times {\rm l}}_{\rm gX}
& = 
b_{\rm g} (b_{\rm X}-b_{\rm e}) \Xi^{(0)}_{0}(\eta,s)
, 
\label{eq: l-l 1}\\
%%%%%%%%%%%%%%%%%%%%
%%%%%%%%%%%%%%%%%%%%
\xi^{{\rm l}\times {\rm v}}_{\rm gX}
& = 
- \left[ b_{\rm g} - \mathcal{B}(\chi)(b_{\rm X}-b_{\rm e}) \right]
\mu \left( s\mathcal{H}f \right) \Xi^{(1)}_{1}(\eta,s)
, \\
%%%%%%%%%%%%%%%%%%%%
%%%%%%%%%%%%%%%%%%%%
\xi^{{v}\times {v}}_{\rm gX}
& = 
\mathcal{B}(\chi)(s\mathcal{H}f)^{2} 
\notag \\
& \quad 
\times \Bigl[ 
\left( \frac{1}{3}-\mu^{2} \right) \Xi^{(2)}_{2}(\eta,s)+ \frac{1}{3}\Xi^{(2)}_{0}(\eta,s)\Bigr]
,
\\
%%%%%%%%%%%%%%%%%%%%
%%%%%%%%%%%%%%%%%%%%
\xi^{{\rm K}\times {\rm l}}_{\rm gX}
& = 
f(b_{\rm X}-b_{\rm e}) \left[ \left( \frac{1}{3}-\mu^{2}\right) \Xi^{(0)}_{2}(\eta,s) +\frac{1}{3}\Xi^{(0)}_{0}(\eta,s)\right]
,
\\
%%%%%%%%%%%%%%%%%%%%
%%%%%%%%%%%%%%%%%%%%
\xi^{{\rm K}\times {v}}_{\rm gX}
& = 
\mathcal{H}sf^{2} \left[ \left( \mu^{3} -\frac{3}{5} \mu \right)\Xi^{(1)}_{3}(\eta,s) - \frac{3}{5} \mu \Xi^{(1)}_{1}(\eta,s)\right]
, \label{eq: K-v 1}
\end{align}
%===========
for the purely local contributions, 
%===========
\begin{align}
\xi^{{\rm nl}\times {\kappa}}_{\rm gX}
& = 
- \mathcal{A}_{\rm X}(\chi) b_{\rm e}\frac{\mathcal{H}}{a^{2}\bar{n}_{\rm e}}
\frac{3\Omega_{\rm m0}\mathcal{H}^{2}_{0}}{\chi}
\int^{\chi}_{0}{\rm d}\chi'\,
a(\chi')\bar{n}_{\rm e}(\chi')
\notag \\
& \times
\left( \chi - \chi' \right) \chi'
\, 
\mathcal{J}\left(\eta', \frac{\chi'}{\chi}s\sqrt{1-\mu^{2}} \right)
, \\
%%%%%%%%%%%%%%%%%%%%
%%%%%%%%%%%%%%%%%%%%
\xi^{\kappa \times \kappa}_{\rm gX}
& = 
\left( \frac{3\Omega_{\rm m0}\mathcal{H}^{2}_{0}}{\chi} \right)^{2}
\int^{\chi}_{0}{\rm d}\chi'\,
\left( \frac{(\chi - \chi') \chi'}{a(\chi')}\right)^{2}\, 
\notag \\
& \times
\mathcal{J}\left(\eta', \frac{\chi'}{\chi}s\sqrt{1-\mu^{2}} \right)
,
\end{align}
%===========
for the purely non-local contributions, and
%===========
\begin{align}
\xi^{{\rm l}\times {\rm nl}}_{\rm gX}
& = 
\Theta(\chi_{2}-\chi_{1})
b_{\rm g}
b_{\rm e}
\mathcal{A}_{\rm X}(\chi)
\mathcal{H}
\, 
\mathcal{J}(\eta, s\sqrt{1-\mu^{2}})
, \\
%%%%%%%%%%%%%%%%%%%%
%%%%%%%%%%%%%%%%%%%%
\xi^{{\rm l}\times {\kappa}}_{\rm gX}
& = 
\left( 
- \Theta(\chi_{2}-\chi_{1})b_{\rm g}
+ \Theta(\chi_{1}-\chi_{2})(b_{\rm X}-b_{\rm e})
\right)
\notag \\
&  \times 
\frac{3\Omega_{\rm m0}\mathcal{H}^{2}_{0}}{a}
s\mu
\, 
\mathcal{J}(\eta, s\sqrt{1-\mu^{2}})
.
\end{align}
%===========
for the cross-correlation between them.

%=================================================
\begin{figure*}
\centering
\includegraphics[width=0.95\textwidth]{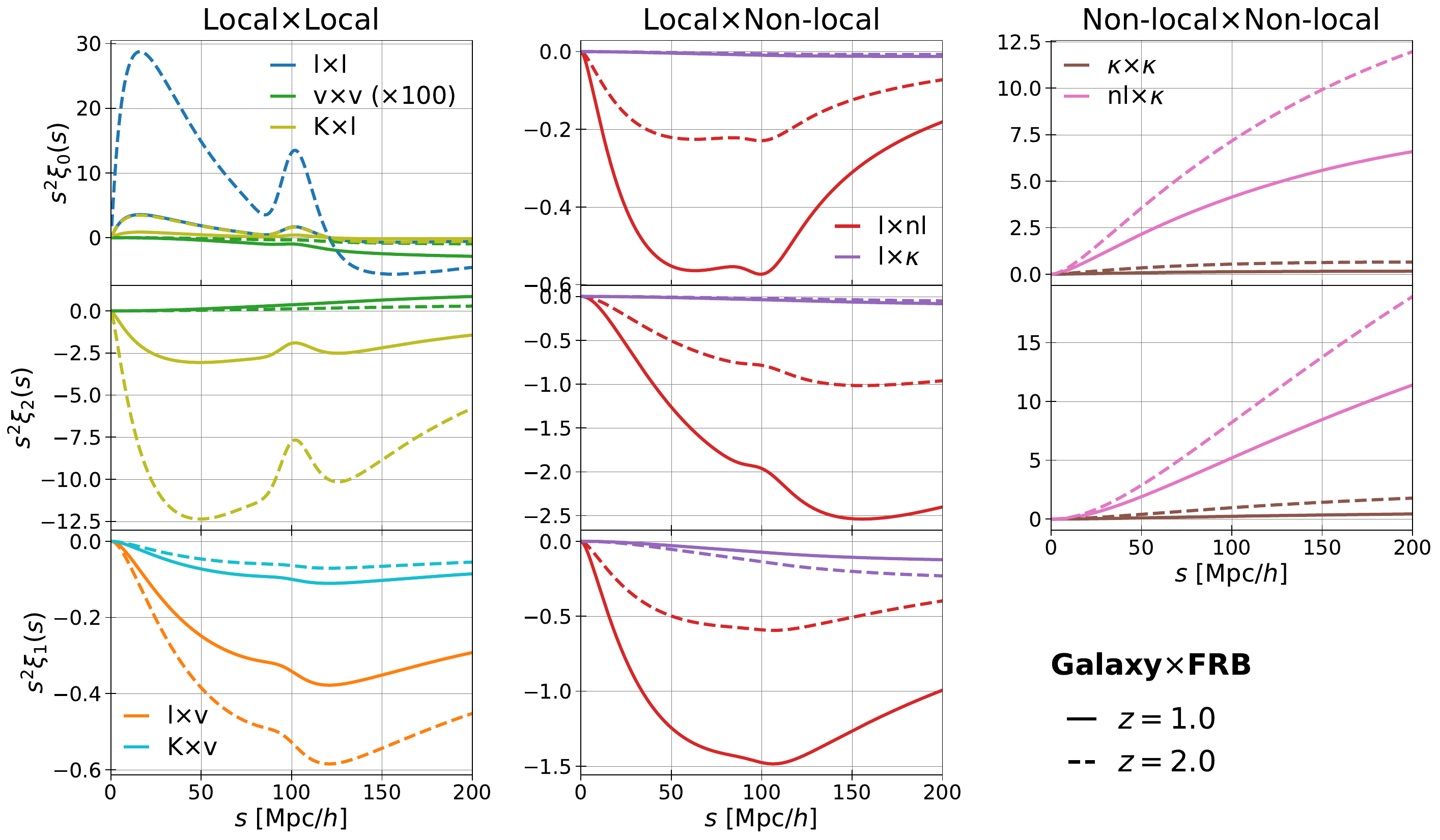}
\caption{Same as Fig.~\ref{fig: FRBs xi multipoles} but for the galaxy-FRB cross-correlations. The result of $\xi^{\rm v\times v}_{\rm gX}$ is multiplied by a factor of 100 for presentation purposes. The bias parameters follow the HI galaxies observed by SKA2, i.e., $b_{\rm SKA}(z) = b_{\rm g} = b_{\rm FRB}$, defined in Eq.~(\ref{eq: bias SKA2}).}
\label{fig: galxFRBs xi multipoles}
\end{figure*}
%=================================================

We quantify the anisotropy in the galaxy-FRB cross-correlation function by performing the multipole expansion (\ref{eq: multipoles}), and present the first three multipoles in Fig.~\ref{fig: galxFRBs xi multipoles}.
We first focus on the even multipoles. There are two dominant contributions: the purely local $\xi^{{\rm l}\times{\rm l}}_{\rm gX}$, and the purely non-local $\xi^{{\rm nl}\times\kappa}_{\rm gX}$, which captures the correlation between magnification bias and the integrated electron density. The absence of the dominant $\xi^{{\rm nl}\times {\rm nl}}_{\rm XY}$ in the FRB-only (see Fig.~\ref{fig: FRBs xi multipoles}) makes it now possible to observe features like the BAO peak in the cross-correlation with galaxies (albeit with a significant non-local contribution). We further find that the contribution from the Kaiser term, which is a unique effect in the galaxy-FRB cross-correlation, is subdominant in the monopole but plays an important role in the quadrupole, especially, at higher redshifts. Turning to the dipole, the primary contributor is $\xi^{{\rm l}\times{\rm nl}}_{\rm gX}$ at $z=1$ and $\xi^{{\rm l}\times{\rm v}}_{\rm gX}$ at $z=2$. The Kaiser term is a minor contribution to the dipole.

%=================================================
\begin{figure}
\centering
\includegraphics[width=0.49\textwidth]{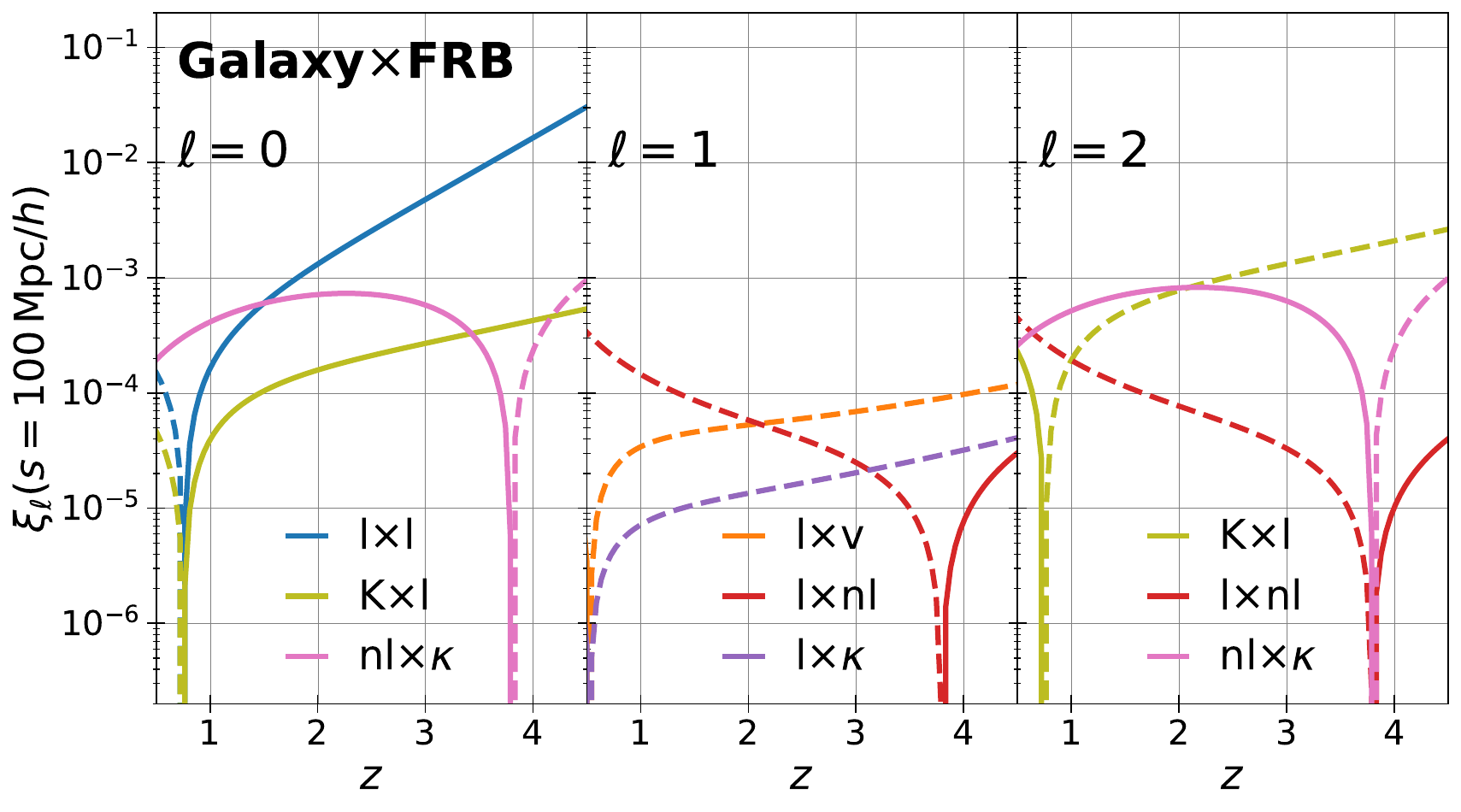}
\caption{Same as Fig.~\ref{fig: FRBs xi zred} but for the galaxy-FRB cross-correlations.}
\label{fig: galxFRB xi zred}
\end{figure}
%=================================================

%=================================================
\begin{figure}
\centering
\includegraphics[width=0.49\textwidth]{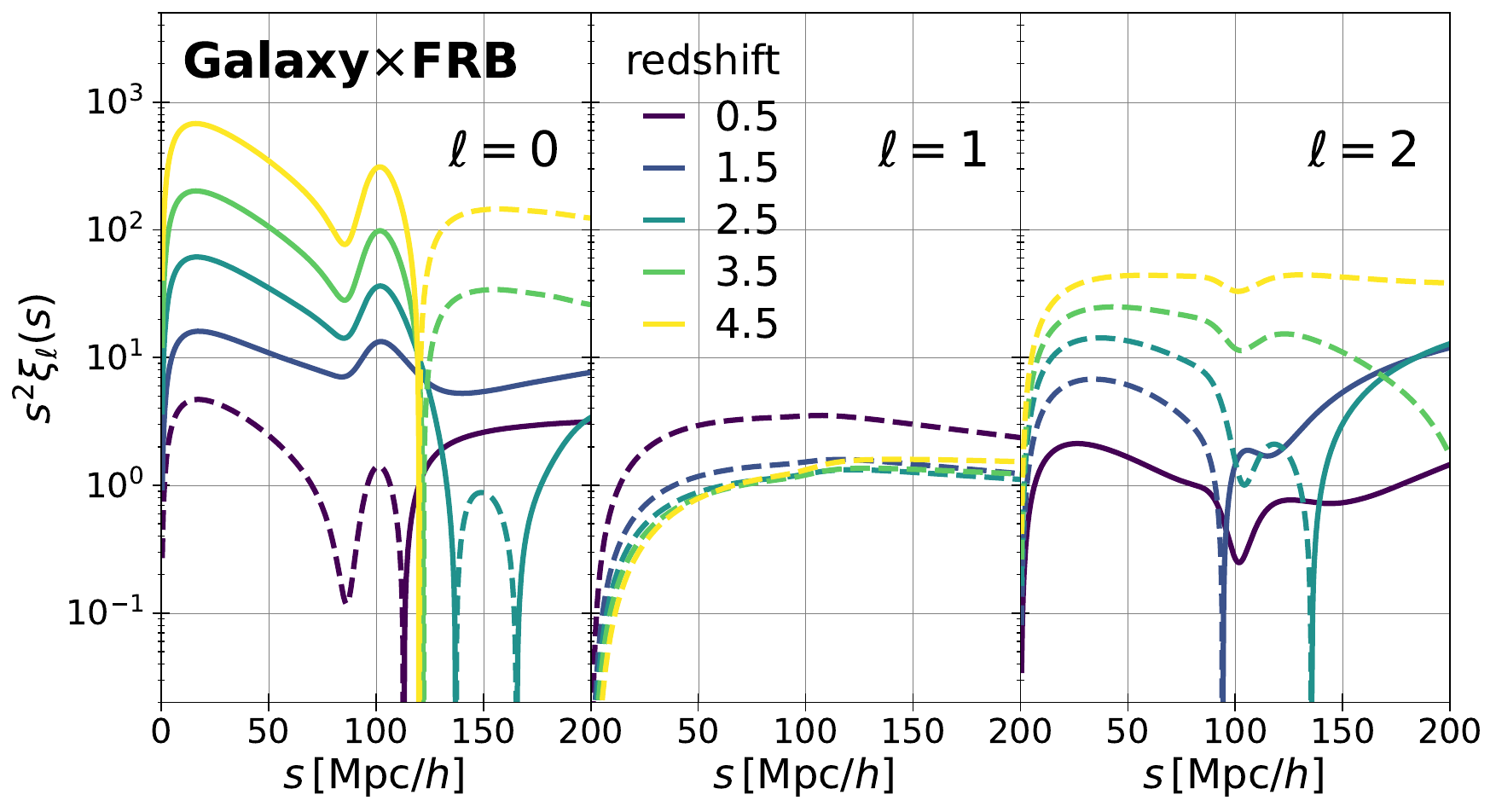}
\caption{Same as Fig.~\ref{fig: FRBs xi multipoles total} but for the galaxy-FRBs cross-correlations.}
\label{fig: galxFRBs xi multipoles total}
\end{figure}
%=================================================

Fig.~\ref{fig: galxFRB xi zred} shows the redshift-dependence of the first three dominant contributions to the multipoles.
Similar to the FRB-only case, shown in Fig.~\ref{fig: FRBs xi zred}, the contributions including the non-local density term vanish around $z\approx 3.8$, where $\mathcal{A}_{\rm X}\approx0$. Since the contributions $(\xi^{{\rm l}\times{\rm l}}_{\rm gX}, \xi^{{\rm K}\times{\rm l}}_{\rm gX})$ and $(\xi^{{\rm l}\times{\rm v}}_{\rm gX}, \xi^{{\rm l}\times \kappa}_{\rm gX})$ are, respectively, proportional to the factor $b_{\rm X}-b_{\rm e}$ and $b_{\rm g} - (b_{\rm X}-b_{\rm e})$, they vanish around $z\approx 0.75$ and $z\approx 0.5$, respectively, in the present setup of the bias parameter~(\ref{eq: bias SKA2}).
The even multipoles are dominated by the local density and Kaiser terms at all redshifts, while the dipole is dominated by the local velocity term at higher redshift and the non-local density term at lower redshift (similar to the FRB-only case). This suggests that it may be possible to isolate and detect the relativistic Doppler contribution from the dipole, assuming that the various astrophysical quantities entering $\mathcal{A}_{\rm X}$ and $\mathcal{B}$ can be determined with sufficient precision.

Finally, we present the behavior of the total signal in Fig.~\ref{fig: galxFRBs xi multipoles total}, which is directly related to the behavior seen in Fig.~\ref{fig: galxFRB xi zred}. At the lowest redshift, the even multipoles show a complex scale dependence due to the coexistence of contributions of different signs, whereas at high redshift, the correlation function multipoles can be explained solely in terms of the $\xi^{{\rm l}\times{\rm l}}_{\rm gX}$ and $\xi^{{\rm K}\times{\rm l}}_{\rm gX}$ terms. The redshift dependence of the dipole does not drastically change compared to the even multipoles due to the two similar contributions, $\xi^{{\rm l}\times{\rm v}}_{\rm gX}$ and $\xi^{{\rm l}\times{\rm nl}}_{\rm gX}$.
High redshift observations of the monopole, dipole, and quadrupole provide us with information about the local density, Doppler effect, and Kaiser effect contributions, suggesting that the three-dimensional clustering in DM space may be a useful cosmological probe, complementary to galaxy redshift surveys.

%%%%%%%%%%%%%%%%%%%%%%%%%%%%%%%%%%%%%%%%%%
%%%%%%%%%%%%%%%%%%%%%%%%%%%%%%%%%%%%%%%%%%
%%%%%%%%%%%%%%%%%%%%%%%%%%%%%%%%%%%%%%%%%%
\section{Future detectability\label{sec: future}}
%%%%%%%%%%%%%%%%%%%%%%%%%%%%%%%%%%%%%%%%%%
%%%%%%%%%%%%%%%%%%%%%%%%%%%%%%%%%%%%%%%%%%
%%%%%%%%%%%%%%%%%%%%%%%%%%%%%%%%%%%%%%%%%%

As demonstrated above, the anisotropic signal in the FRB correlation function in DM space may provide a complementary method to test our cosmological model and theory of gravity. In this section, we perform a Fisher matrix analysis to investigate the future detectability of this signal, assuming an SKA-like survey specification, assuming the detection of a large number of FRBs. To this end, it is more convenient to work with two-point statistics in Fourier space than in real space. Thus, we first define the multipole power spectrum in Sec.~\ref{sec: power spectrum}, and investigate their detectability in Sec.~\ref{sec: SN}.

%%%%%%%%%%%%%%%%%%%%%%%%%%%%%%%%%%%%%%%%%%
%%%%%%%%%%%%%%%%%%%%%%%%%%%%%%%%%%%%%%%%%%
\subsection{Power spectrum multipoles\label{sec: power spectrum}}
%%%%%%%%%%%%%%%%%%%%%%%%%%%%%%%%%%%%%%%%%%
%%%%%%%%%%%%%%%%%%%%%%%%%%%%%%%%%%%%%%%%%%

The power spectrum multipoles are related to the multipoles of the correlation function via:
%===========
\begin{align}
P_{\ell}(k) = 4\pi(-i)^{\ell}\int{\rm d}s\, s^{2} j_{\ell}(ks) \xi_{\ell}(s) . \label{eq: def P ell}
\end{align}
%===========
where we still take the distant-observer limit. To compute the multipole power spectrum, we perform this integral numerically.

%=================================================
\begin{figure}
\centering
\includegraphics[width=0.49\textwidth]{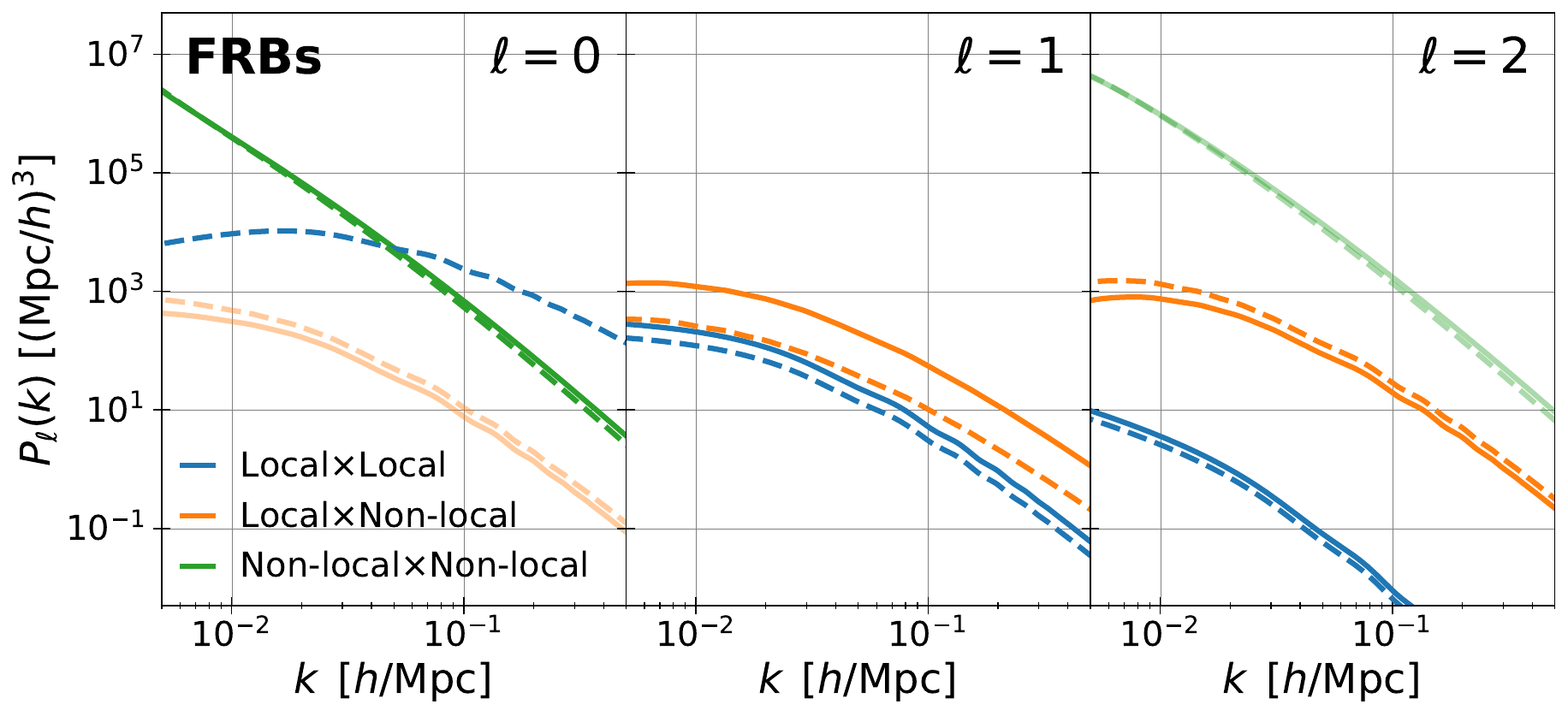}
\includegraphics[width=0.49\textwidth]{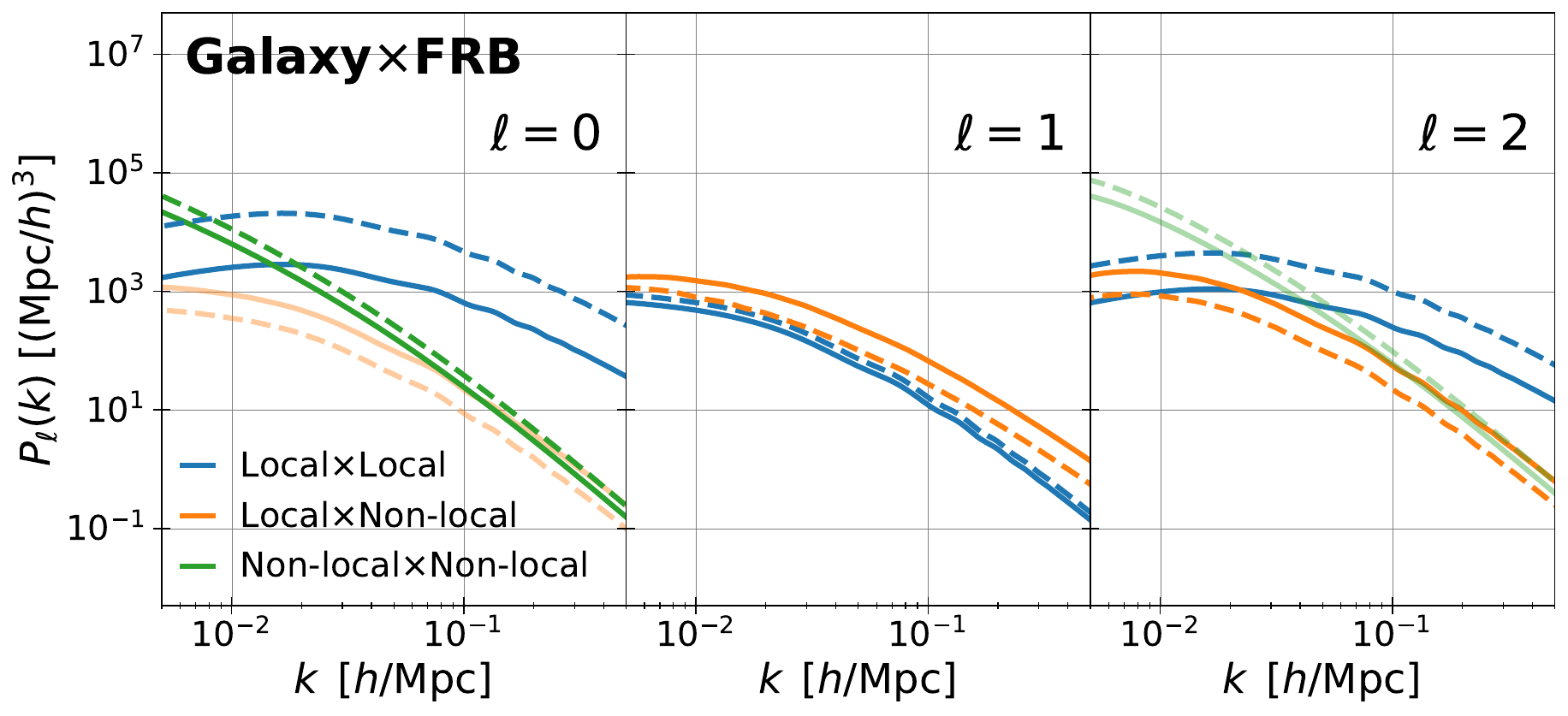}
\caption{Power spectrum multipoles for the FRB power spectra (top panels) and galaxy-FRB cross-power spectra (bottom panels) at $z=1$ (solid lines) and at $z=2$ (dashed lines). The thin lines indicate a negative amplitude. Blue, orange, and green lines, respectively, show the sum of the pure local contributions, crosstalk between the local and non-local contributions, and pure non-local contributions.}
\label{fig: FRBs P multipoles total}
\end{figure}
%=================================================

In Fig.~\ref{fig: FRBs P multipoles total}, we present the multipole power spectra at $z=1$ and $2$. These results are simply the Fourier counterpart of the correlation function results shown in Figs.~\ref{fig: FRBs xi multipoles} and \ref{fig: galxFRBs xi multipoles}. We see again that the even multipoles of the FRB power spectra (top panels) are dominated by the pure non-local contribution whereas the dipole is not because of the absence of the non-local density contribution. Likewise, the purely non-local contribution to the galaxy-FRB cross-power spectra (bottom panels) is suppressed compared to the FRB-only case. The behavior of the monopole in the FRB power spectra is consistent with the angular power spectrum predictions found in Ref.~\cite{2021PhRvD.103l3544A}.

%%%%%%%%%%%%%%%%%%%%%%%%%%%%%%%%%%%%%%%%%%
%%%%%%%%%%%%%%%%%%%%%%%%%%%%%%%%%%%%%%%%%%
\subsection{Signal-to-noise ratio\label{sec: SN}}
%%%%%%%%%%%%%%%%%%%%%%%%%%%%%%%%%%%%%%%%%%
%%%%%%%%%%%%%%%%%%%%%%%%%%%%%%%%%%%%%%%%%%

In this subsection, we discuss the future detectability of DM-space anisotropies. To estimate the signal-to-noise ratio, we compute the covariance matrix of the power spectrum multipoles between the objects X and Y by neglecting the non-Gaussian contribution~(e.g., Refs.~\cite{2003ApJ...595..577Y,2006PASJ...58...93Y,2010PhRvD..82f3522T,2021JCAP...01..036N})
%===========
\begin{align}
{\rm COV}_{\ell}(k,k') &\equiv \Braket{\Delta P^{\rm XY}_{\ell}(k) \left(\Delta P^{\rm XY}_{\ell}(k') \right)^{*}} \\
&\equiv
\frac{\delta_{\rm D}(k-k')}{k^{2}}
\frac{2\pi^{2}}{V}
\sigma^{2}_{\ell}(k) ,
\end{align}
where, we have defined the variance $\sigma^{2}_{\ell}(k)$ as
%===========
\begin{align}
\sigma^{2}_{\ell}(k)
& = 
\frac{1}{2} \sum_{\ell_{1},\ell_{2}} 
\Biggl[
\left( P^{\rm XX}_{\ell_{1}}(k) + \frac{\delta_{\ell_{1},0}}{n_{\rm X}} \right)
\notag \\
& \times
\left( P^{\rm YY}_{\ell_{2}}(k) + \frac{\delta_{\ell_{2},0}}{n_{\rm Y}} \right)
+ (-1)^{\ell} P^{\rm XY}_{\ell_{1}}(k) P^{\rm YX}_{\ell_{2}}(k) 
\Biggr] \notag \\
& \times 
\left( \frac{2\ell+1}{2} \right)^{2}
\int{\rm d}\mu\,
\mathcal{L}_{\ell}(\mu)\mathcal{L}_{\ell}(\mu)\mathcal{L}_{\ell_{1}}(\mu)\mathcal{L}_{\ell_{2}}(-\mu) . \label{eq: sigma ell}
\end{align}
%===========
Here the quantities $n_{\rm X/Y}$ and $V$ stand for the number density of the ${\rm X}/{\rm Y}$ samples, and the observed volume, respectively. We numerically compute the multipole power spectrum by Eq.~(\ref{eq: def P ell}).

In the signal-to-noise ratio analysis, we consider two cases: FRB cross-power spectrum and galaxy-FRB cross-power spectrum.
As we expect to detect roughly $10^{5}$--$10^{6}$ FRBs in the SKA era~\cite{2020MNRAS.497.4107H}, we set the number of FRBs of each species to $10^{4}$--$10^{6}$. We use the galaxy number density given by the specifications of SKA2 HI galaxies in Eq.~(B1) in Ref.~\cite{2016ApJ...817...26B}:
%===========
\begin{align}
\frac{{\rm d}N}{{\rm d}z} = 10^{c_{1}}z^{c_{2}}\exp\left( -c_{3}z \right) \quad {\rm [deg^{-2}]}, 
\label{eq: number of galaxies SKA2}
\end{align}
%===========
with $c_{1} = 6.319$, $c_{2} = 1.736$, and $c_{3} = 5.424$. The average number of galaxies per redshift interval can be related to the number density per volume by $n = {\rm d}N/{\rm d}z \left( \Delta z f_{\rm sky} /V \right)$ with $\Delta z = 0.1$ and $f_{\rm sky} = 30000\,{\rm deg}^{2}$ being the width of redshift bins and the fractional sky coverage, respectively. We compute the survey volume as $V = 4\pi/3\, \left(\chi(z+\Delta z/2)^{3} - \chi(z-\Delta z/2)^{3}\right)$. Note that, optmistically, we assume that all the detected FRBs are located within the same $\Delta z=0.1$ interval. In computing the galaxy auto-power spectrum entering the covariance, we ignore the negligible small contribution from relativistic  effects~\cite{2017PhRvD..95d3530H,2018JCAP...05..043L,2022MNRAS.511.2732S}, and simply consider the standard Kaiser effect. We then estimate the signal-to-noise ratio by
%===========
\begin{align}
\left( \frac{\rm S}{\rm N} \right)^{2}
&= \int^{k_{\rm max}}_{k_{\rm min}} {\rm d}k\,  P^{\rm XY}_{\ell}(k)\left[ {\rm COV}_{\ell}(k,k') \right]^{-1}\left( P^{\rm XY}_{\ell}(k') \right)^{*} \notag \\
&= V\int^{k_{\rm max}}_{k_{\rm min}} \frac{{\rm d}k}{2\pi^{2}} k^{2} \frac{\left| P^{\rm XY}_{\ell}(k) \right|^{2}}{\sigma^{2}_{\ell}(k)} , \label{eq: SN}
\end{align}
%===========
where we set $k_{\rm min} = 2\pi V^{-1/3}$ and $k_{\rm max} = 0.1\, h/{\rm Mpc}$ to avoid the non-linear contribution to the density perturbations (see Appendix~\ref{sec: NL dipole} for the non-linear contributions to the dipole).

%=================================================
\begin{figure*}
\centering
\includegraphics[width=0.99\textwidth]{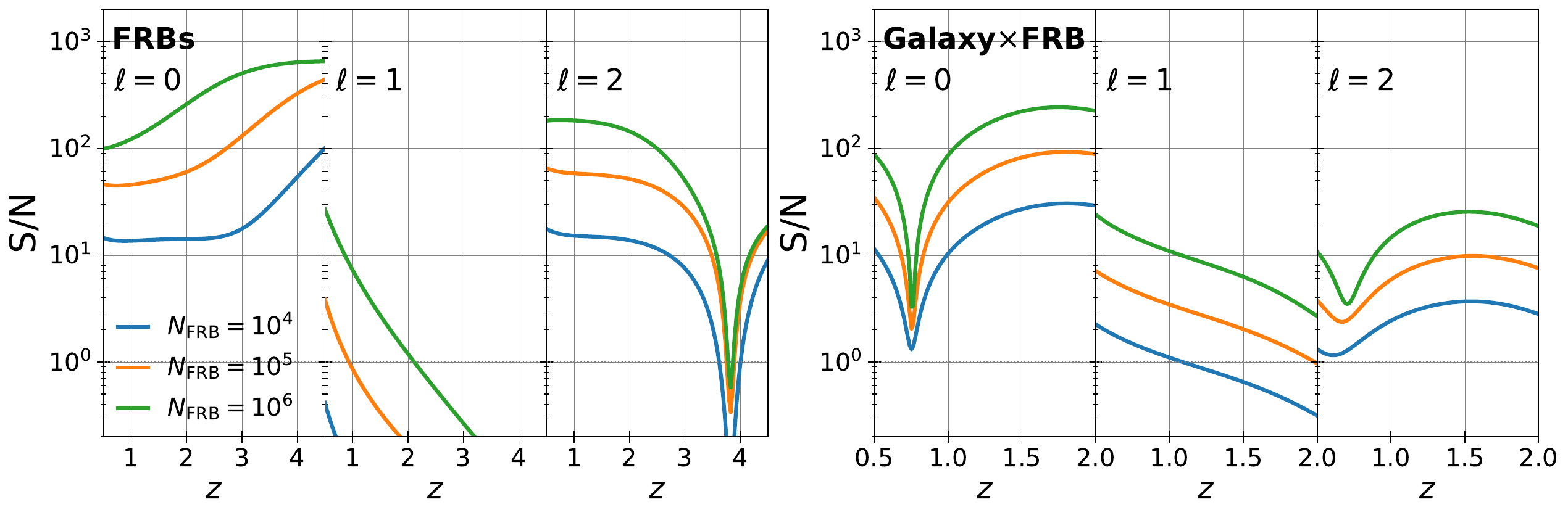}
\caption{Signal-to-noise ratio of the multipole power spectra in the SKA-like survey specifications for FRBs power spectra (left panels) and for galaxy-FRB cross-power spectra (right panels), respectively. The blue, orange, and green lines represent, respectively, the results with the number of FRBs, $N_{\rm FRB} = 10^{4}$, $10^{5}$, and $10^{6}$.}
\label{fig: FRBs P SN}
\end{figure*}
%=================================================

Using Eq.~(\ref{eq: SN}), we compute the redshift dependence of the signal-to-noise ratio of the multipole power spectra assuming the SKA-like survey, shown in Fig.~\ref{fig: FRBs P SN}. We found that the even multipole power spectra could be detected with the high statistical significance, ${\rm S}/{\rm N} \approx$ $10^{1}$--$10^{2}$. 
Interestingly, we observe a suppression of the signal-to-noise ratio at $z\approx 3.8$ in the quadrupole of the FRB alone result. This is because the quadrupole is dominated by the non-local density term, which becomes zero around $z\approx 3.8$ (see Fig.~\ref{fig: FRBs xi multipoles total}). 
Overall, the signal-to-noise ratio for the even multipoles in the FRB-only case is larger than that of the cross-power spectra case, mostly due to the purely non-local density term $\xi^{{\rm nl}\times{\rm nl}}_{\rm XY}$, which significantly amplifies the even multipole signal.

Turning to the signal-to-noise ratio for the dipole, observations at low redshift have the best chances of detecting the dipole, although in this case the dipole is dominated by not the relativistic effect but the non-local density contribution. However, assuming a more optimistic number of FRBs $\sim 10^{6}$, the chance for detecting the relativistic effect through the dipole improves more, with ${\rm S}/{\rm N} \gtrsim 1$ at $z\approx 2$. Moreover, cross-correlations between FRBs and galaxies, the signal-to-noise ratio of the dipole improves, particularly ${\rm S}/{\rm N} \approx 10$ for the most optimistic case with $N_{\rm FRB}=10^{6}$.
Hence, observing a large number of FRBs and making a three-dimensional map of FRBs may be an interesting new approach to the detection of relativistic effects. In addition, the even multipole anisotropy in DM space contains a wealth of cosmological information, which would also provide complementary information to galaxy redshift surveys.

%%%%%%%%%%%%%%%%%%%%%%%%%%%%%%%%%%%%%%%%%%
%%%%%%%%%%%%%%%%%%%%%%%%%%%%%%%%%%%%%%%%%%
%%%%%%%%%%%%%%%%%%%%%%%%%%%%%%%%%%%%%%%%%%
\section{Summary\label{sec: summary}}
%%%%%%%%%%%%%%%%%%%%%%%%%%%%%%%%%%%%%%%%%%
%%%%%%%%%%%%%%%%%%%%%%%%%%%%%%%%%%%%%%%%%%
%%%%%%%%%%%%%%%%%%%%%%%%%%%%%%%%%%%%%%%%%%

We have investigated the three-dimensional clustering of the sources emitting electromagnetic pulses such as fast radio bursts (FRBs) in dispersion measure (DM) space. The DM of pulses can be exploited as a cosmological distance measure, although it is systematically affected by inhomogeneities in the electron density and special and general relativistic effects, as is the redshift measured in galaxy redshift surveys. Accordingly, the observed DM-based clustering is affected by DM space distortions.
Following the footsteps of Ref.~\cite{2014ApJ...780L..33M,2015PhRvL.115l1301M,2019PhRvD.100h3533K,2020PhRvD.102b3528R}, we formulate the two-point statistics in DM space including all the possible dominant contributions to the DM.

The observed anisotropy in the correlation function or power spectrum is induced by the contributions from the non-local integral term along the line of sight and from the Doppler term, which is a major relativistic contribution to the DM. Performing a multipole expansion, we found that the even multipoles are primarily dominated by both the non-local density and local density contributions, whereas the dipole moment receives a contribution from the Doppler term, suggesting that observing three-dimensional clustering may isolate this relativistic effect. We note that, as in the case of redshift space distortions, the nonvanishing dipole appears only when we correlate two different biased objects, i.e., cross correlation between two subsamples of FRBs with different biases or galaxy-FRB cross correlation.

Based on the derived analytical model for the correlation function or power spectrum in DM space, we further investigate the future detectability by performing the Fisher matrix analysis. Assuming an SKA-like survey, the signal-to-noise ratio for the even multipoles reaches ${\rm S}/{\rm N} \approx 100$ and that the dipole may be detectable if a sufficiently large number of FRBs are measured. In the optimistic case where the observed number of FRBs is $10^{5}$--$10^{6}$, the signal-to-noise ratio exceeds unity even at high redshift, where the Doppler term dominates the dipole, and hence the high-redshift measurement would be an interesting probe to test gravity, complementary to galaxy redshift surveys.

In the Fisher matrix analysis, we have made several simplifications to make the problem more tractable. Importantly, we have ignored the local contribution to the DM from the electron density of host galaxies. This contribution would lead to the suppression of the correlation signal at small scales if its contribution is a random quantity uncorrelated to the large-scale structure~\cite{2021JCAP...01..036N}, effectively increasing the uncertainty in the measurement. This assumption could be refined, for example, through the use of hydrodynamic simulations~(e.g., Refs.~\cite{2014MNRAS.444.1518V,2019MNRAS.484.1637J,2021MNRAS.502.2615T}), providing a more realistic forecast. We have also assumed a relatively high number density of FRBs, and thus the results presented depend on the ability of future experiments, such as the SKA, to achieve this detection rate. Finally, we have set the bias parameter of FRBs to be similar to that of the HI galaxies observed by SKA2. Furthermore, we have ignored the gravitational potential contribution to the DM as it is expected to be negligible at large scales. However, if FRBs reside in the deep potential well of dark matter haloes, the gravitational potential contribution would play a certain role at small scales~(see redshift-space examples \cite{2019MNRAS.483.2671B,2020MNRAS.498..981S,2022MNRAS.511.2732S,2023MNRAS.524.4472S}). Properly taking into account these contributions, the anisotropy of the three-dimensional clustering in DM space could become a more promising and robust probe for testing gravity on cosmological scales in the next decades.

%%%%%%%%%%%%%%%%%%%%%%%%%%%%%%%%%%%%%%%%%%
\begin{acknowledgments}
We would like to thank Francesca Lepori and Ruth Durrer for interesting discussions. SS is supported by JSPS Overseas Research Fellowships. SS thanks Julien Devriendt for being a host during the Balzan program. This work was initiated during the visiting program supported by a Centre for Cosmological Studies Balzan Fellowship.
\end{acknowledgments}
%%%%%%%%%%%%%%%%%%%%%%%%%%%%%%%%%%%%%%%%%%

%%%%%%%%%%%%%%%%%%%%%%%%%%%%%%%%%%%%%%%%%%
\appendix
%%%%%%%%%%%%%%%%%%%%%%%%%%%%%%%%%%%%%%%%%%

%%%%%%%%%%%%%%%%%%%%%%%%%%%%%%%%%%%%%%%%%%
%%%%%%%%%%%%%%%%%%%%%%%%%%%%%%%%%%%%%%%%%%
%%%%%%%%%%%%%%%%%%%%%%%%%%%%%%%%%%%%%%%%%%
\section{Wide-angle correction\label{sec: app wa}}
%%%%%%%%%%%%%%%%%%%%%%%%%%%%%%%%%%%%%%%%%%
%%%%%%%%%%%%%%%%%%%%%%%%%%%%%%%%%%%%%%%%%%
%%%%%%%%%%%%%%%%%%%%%%%%%%%%%%%%%%%%%%%%%%

Here we derive the wide-angle correction to the local contributions. Since the wide-angle correction to the non-local contribution would be negligible for the multipole analysis~\cite{2018JCAP...10..032T,2018JCAP...03..019T}, we focus only on the local contributions here.

In general, the correlation function is given as a function of two vectors $\bm{x}_{1}$ and $\bm{x}_{2}$, but the correlation function in the distant observer limit can be given as a function of the separation $s = |\bm{x}_{2} - \bm{x}_{1}|$ and the directional cosine between the separation vector and the fixed line-of-sight vector: $\mu = \hat{\bm{n}}\cdot\hat{\bm{z}}$. However, when we do not take the distant observer limit, the correlation function can be characterized by three variables: the separation $s$, line-of-sight distance specifically pointing to a midpoint $\chi = |\bm{d}| = |\bm{x}_{1} + \bm{x}_{2}|/2$, and directional cosine between the separation vector and the line-of-sight vector $\mu = \hat{\bm{n}}\cdot\hat{\bm{d}}$, i.e., $\xi(\bm{x}_{1},\bm{x}_{2}) = \xi(x,\chi,\mu)$.

We then expand the correlation function in powers of $(s/\chi)$ to split the $\chi$ dependence from the correlation function as follows:
%===========
\begin{align}
\xi(s,\chi,\mu) &=
\xi^{(\rm pp)} (s,\mu) + \xi^{(\rm wa)} (s,\mu) + O\left( \left( \frac{s}{\chi} \right)^{2} \right) , \label{eq: wa expansion app}
\end{align}
%===========
where the first and second terms on the right-hand side are, respectively, the expression in the distant observer limit ($s/\chi\to 0$), and the leading-order correction to the wide-angle effect $\propto  \left( s/\chi \right)^{1}$.

\begin{figure}
\centering
\includegraphics[width=0.49\textwidth]{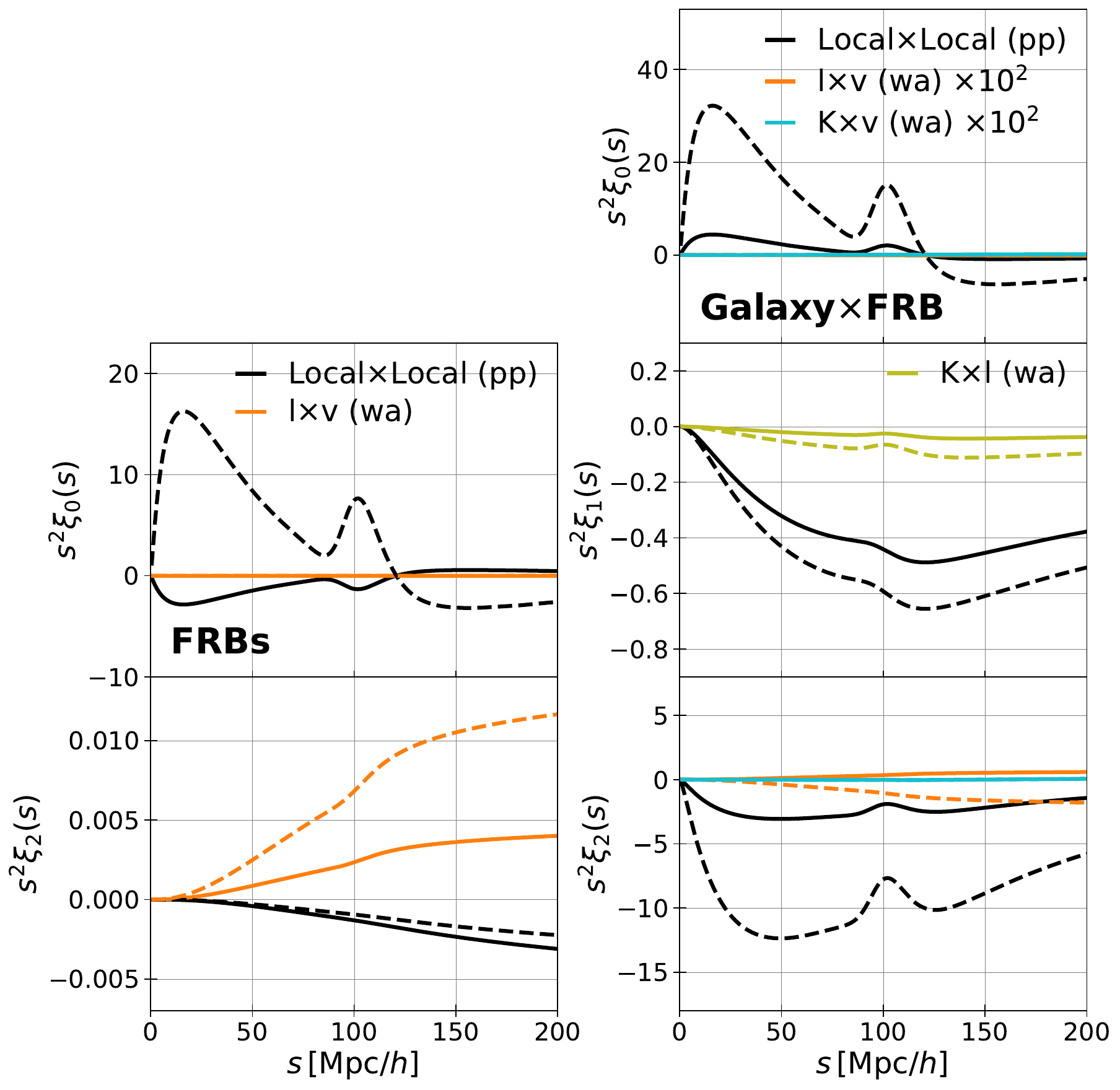}
\caption{Wide-angle corrections to the multipoles of the FRBs cross-correlation function at $z=1.0$ (solid lines) and at $z=2.0$ (dashed lines). The black lines present the total local contributions, i.e., the sum of Eqs.~(\ref{eq: l-l})--(\ref{eq: v-v}) for the FRB alone case and the sum of Eqs.~(\ref{eq: l-l 1})--(\ref{eq: K-v 1}) for the galaxy-FRB cross-correlation case. For presentation purposes, the results in the right panels for $\xi^{{\rm l}\times{\rm v}, {\rm (wa)}}_{\rm gX}$ and $\xi^{{\rm K}\times{\rm v}, {\rm (wa)}}_{\rm gX}$ are multiplied by $10^{2}$.}
\label{fig: xi wa}
\end{figure}

First, we look at the wide-angle correction to the FRB cross-correlation. Among the local contributions, only the cross-correlation between $\delta^{\rm l}$ and $\delta^{\rm v}$ has a nonvanishing wide-angle correction $O(s/\chi)$, which is explicitly given by
%===========
\begin{align}
\xi^{\rm {\rm l}\times {\rm v}, (\rm wa)}_{\rm XY} & = 
\left( \frac{s}{\chi}\right)
(b_{\rm X} + b_{\rm Y} - 2b_{\rm e})
\notag \\
& \quad \times 
\left( \mu^{2} - \frac{1}{2}\right)\left( s\mathcal{H}f \right) \Xi^{(1)}_{1}(\eta,s)
.
\end{align}
%===========
Next, for the galaxy-FRBs cross-correlation case, the nonvanishing wide-angle corrections at the leading order are given by
%===========
\begin{align}
\xi^{\rm {\rm K}\times {\rm l}, (\rm wa)}_{\rm gX} & = 
\left( \frac{s}{\chi}\right)
f(b_{\rm X}-b_{\rm e})(1-2\mu^{2})\mu \Xi^{(0)}_{2}(\eta, s)
, \\
\xi^{\rm {\rm K}\times {\rm v}, (\rm wa)}_{\rm gX} & = 
\left( \frac{s}{\chi}\right)
s\mathcal{H}f^{2}\left(\mu^{2}-\frac{1}{2}\right)
\notag \\
& \quad \times 
\left[ \mu^{2} \Xi^{(1)}_{3}(\eta,s)- \Xi^{(2)}_{2}(\eta,s)\right]
.
\end{align}
%===========
Using the derived analytical expressions, we compare the wide-angle correction to the total local contributions in Fig.~\ref{fig: xi wa}. Clearly, the wide-angle contribution to the galaxy-FRB cross-correlation is negligibly small compared to the result of the plane-parallel limit. On the other hand, the wide-angle correction to the quadrupole of the FRB cross-correlation exhibits a non-negligible impact. However, the total quadrupole signal is mostly dominated by the non-local contributions, as far as we are interested in the total signal, the wide-angle correction can be safely ignored.

%=================================================
%=================================================

%%%%%%%%%%%%%%%%%%%%%%%%%%%%%%%%%%%%%%%%%%
%%%%%%%%%%%%%%%%%%%%%%%%%%%%%%%%%%%%%%%%%%
%%%%%%%%%%%%%%%%%%%%%%%%%%%%%%%%%%%%%%%%%%
\section{Nonlinear impact on the dipole signal\label{sec: NL dipole}}
%%%%%%%%%%%%%%%%%%%%%%%%%%%%%%%%%%%%%%%%%%
%%%%%%%%%%%%%%%%%%%%%%%%%%%%%%%%%%%%%%%%%%
%%%%%%%%%%%%%%%%%%%%%%%%%%%%%%%%%%%%%%%%%%

%=================================================
\begin{figure}
\centering
\includegraphics[width=0.49\textwidth]{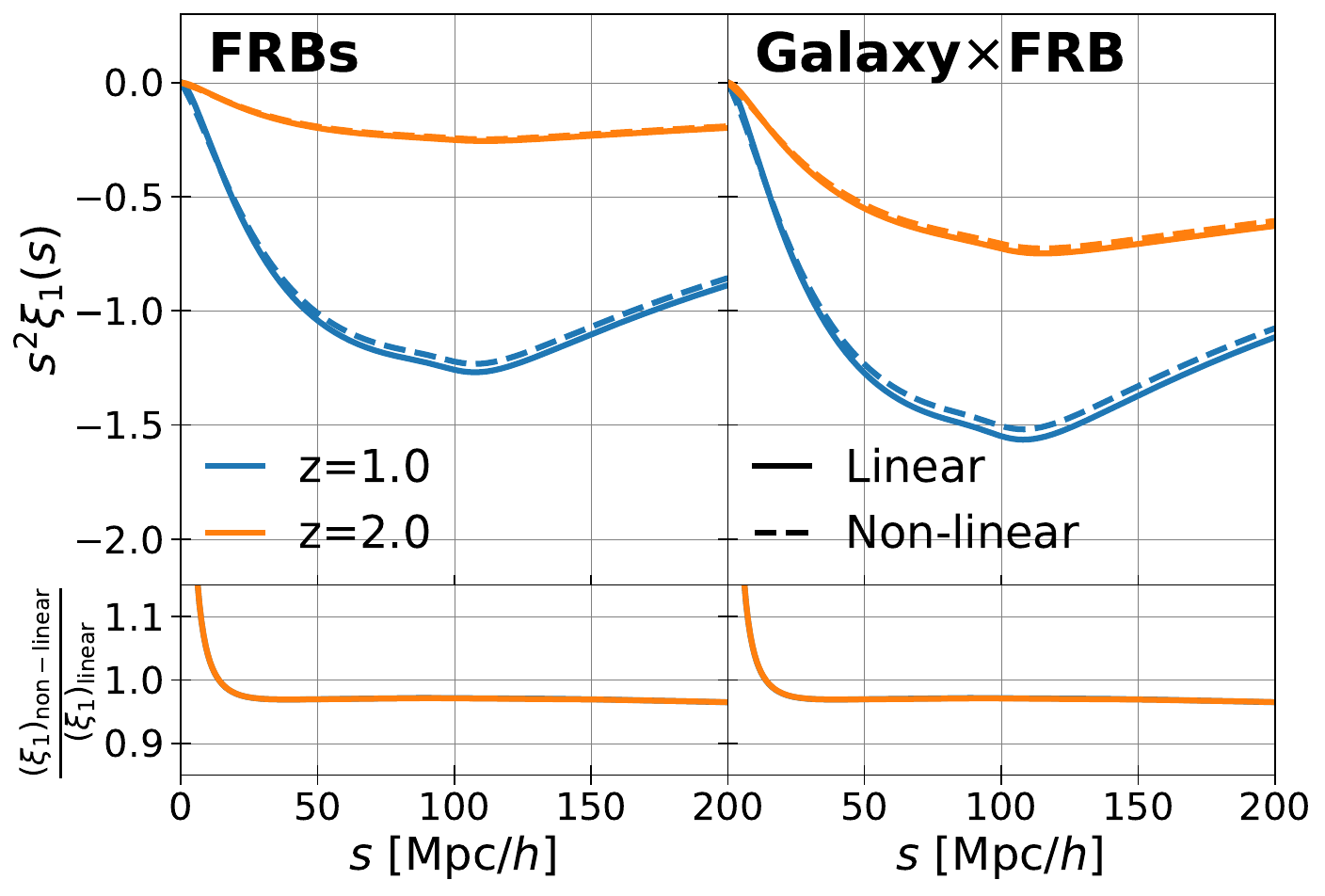}
\caption{Dipole from the cross correlation between the local and non-local terms at $z=1.0$ (blue) and $z=2.0$ (orange). The solid and dashed lines represent, respectively, the prediction with the linear matter power spectrum and with the non-linear matter power spectrum. The bottom panels show the ratio between the predictions with the linear and non-linear matter power spectrum. We note that the blue and orange lines in the bottom panels mostly overlap, and only the orange lines are visible.}
\label{fig: xi nl}
\end{figure}
%=================================================

As we have shown in Figs.~\ref{fig: FRBs xi multipoles} and \ref{fig: galxFRBs xi multipoles}, the cross-correlation between the local and non-local terms is the dominant contributor to the dipole signal. In this Appendix, we discuss the impact of the non-linear density growth on the dipole signal, particularly focusing on the local and non-local crosstalk.

To include the non-linear density growth into the prediction, we simply replace the linear matter power spectrum in $\Xi^{(n)}_{\ell}$ and $\mathcal{J}$, respectively given in Eqs.~(\ref{eq: def Xi}) and (\ref{eq: def calJ}), by the non-linear matter power spectrum computed by using \texttt{CLASS} with a non-linear output option \texttt{HALOFIT}~\cite{2011arXiv1104.2932L,2011JCAP...10..037A}.
Fig.~\ref{fig: xi nl} show the impact of the non-linear matter power spectrum on the dipole signal. As we clearly see that, as long as we restrict our analysis to the large-scale signal, we safely ignore the non-linear matter growth effect because it suppresses the dipole signal only a few percent level at $s\gtrsim 20\, {\rm Mpc}/h$.

%%%%%%%%%%%%%%%%%%%%%%%%%%%%%%%%%%%%%%%%%%
\bibliography{DM_space}% Produces the bibliography via BibTeX.
\end{document}